\documentclass[12pt]{article}
\usepackage{amsmath,amssymb,epsfig}
\renewcommand{\theequation}{\thesection.\arabic{equation}}

\newlength{\extraspace}
\newlength{\extraspaces}
\newcommand{\be}{\begin{equation}
\addtolength{\abovedisplayskip}{\extraspaces}
\addtolength{\belowdisplayskip}{\extraspaces}
\addtolength{\abovedisplayshortskip}{\extraspace}
\addtolength{\belowdisplayshortskip}{\extraspace}}
\newcommand{\ee}{\end{equation}}
 
\newcommand{\ba}{\begin{eqnarray}
\addtolength{\abovedisplayskip}{\extraspaces}
\addtolength{\belowdisplayskip}{\extraspaces}
\addtolength{\abovedisplayshortskip}{\extraspace}
\addtolength{\belowdisplayshortskip}{\extraspace}}
\newcommand{\ea}{\end{eqnarray}}

\newcommand{\bas}{\begin{eqnarray*}
\addtolength{\abovedisplayskip}{\extraspaces}
\addtolength{\belowdisplayskip}{\extraspaces}
\addtolength{\abovedisplayshortskip}{\extraspace}
\addtolength{\belowdisplayshortskip}{\extraspace}}
\newcommand{\eas}{\end{eqnarray*}}
 
\newcounter{subequation}[equation]
\makeatletter

\expandafter\let\expandafter
\reset@font\csname reset@font\endcsname

\def\subeqnarray{\arraycolsep1pt
    \def\@eqnnum\stepcounter##1{\stepcounter{subequation}%
        {\reset@font\rm(\theequation\alph{subequation})}}
\jot5mm     \eqnarray}

\def\subarray{\arraycolsep1pt
    \def\@eqnnum\stepcounter##1{\stepcounter{subequation}%
        {\reset@font\rm(\alph{subequation})}}
\jot5mm     \eqnarray}

\makeatother

\newcommand{\newsection}[1]{
\vspace{15mm}
\pagebreak[3]
\addtocounter{section}{1}
\setcounter{equation}{0}
\setcounter{subsection}{0}

\setcounter{footnote}{0}
\addcontentsline{toc}{section}
{\protect\numberline{\arabic{section}}{#1}}
 
\begin{flushleft}
{\large\bf \thesection. #1}
\end{flushleft}
\nopagebreak
\medskip
\nopagebreak}
 
\newcommand{\newsubsection}[1]{
\vspace{1cm}
\pagebreak[3]
\addtocounter{subsection}{1}

\addcontentsline{toc}{subsection}
{\protect\numberline{\thesection.\arabic{subsection}}{#1}}
 
\noindent{ \bf \thesection.\arabic{subsection} #1}
\nopagebreak
\vspace{2mm}
\nopagebreak}

\newcommand{\newappendix}[1]{
\vspace{15mm}
\pagebreak[3]
\addtocounter{section}{1}
\setcounter{equation}{0}
\setcounter{subsection}{0}

\addcontentsline{toc}{section}
{\protect\numberline{\thesection}{#1}}

\renewcommand{\theequation}{\Alph{section}.\arabic{equation}}
\begin{flushleft}
{\large\bf \Alph{section}: #1}
\end{flushleft}
\nopagebreak
\medskip
\nopagebreak}

\newcommand{\N}{\mathbb{N}}
\newcommand{\C}{\mathbb{C}}

\newcommand{\R}{\mathbb{R}}

\renewcommand{\P}{\mathbb{P}}

\newcommand{\bra}{\langle}
\newcommand{\ket}{\rangle}
\newcommand{\ra}{\rightarrow}

\newcommand{\rra}{\ \longrightarrow \ }

\newcommand{\is}{ &\! =\! & }
\newcommand{\nonum}{\nonumber \\[1.5mm]}
\newcommand{\sspace}{\makebox[1cm]{ }}
\newcommand{\bspace}{\makebox[2cm]{ }}
\newcommand{\nspace}{\!\!\!\!\!\!\!\!\!\!}

\newcommand{\Tr}{{\rm Tr}}

\renewcommand{\th}{{\theta}}
\newcommand{\eps}{\epsilon}
\newcommand{\lb}{\lambda}
\newcommand{\om}{\omega}

\newcommand{\dd}{{\partial}}

\newcommand{\cA}{{\cal A}}

\newcommand{\cD}{{\cal D}}

\newcommand{\cO}{{\cal O}}

\newcommand{\vp}{\varphi} 
\newcommand{\bP}{{\bf P}}
\newcommand{\bH}{{\bf H}}
\newcommand{\bL}{{\bf L}}

\begin{document}
\mbox{} 
\vspace{1cm}

\begin{center}
\mbox{{\large \bf Temporal breakdown and Borel resummation}}\\[3mm]
\mbox{{\large \bf in the complex Langevin method}} 
\vspace{2.5cm}

{{\sc A.~Duncan\footnote{tony@dectony.phyast.pitt.edu}
 and M. Niedermaier\footnote{mnie@pitt.edu} 
}}
\\[4mm]
\vspace{7mm} 
{\small \sl Department of Physics and Astronomy}\\
{\small \sl University of Pittsburgh}\\
{\small \sl 100 Allen Hall}\\
{\small \sl Pittsburgh, PA 15260, USA} 
\vspace{15mm}
\begin{quote} 
We reexamine the Parisi-Klauder conjecture for complex $e^{i\th/2} \phi^4$ 
measures with a Wick rotation angle $0 \leq \th/2 \leq \pi/2$ interpolating 
between Euclidean and Lorentzian signature. Our main result is 
that the asymptotics for short stochastic times $t$  
encapsulates information also about the equilibrium aspects. 
The moments evaluated with the complex measure and with the real 
measure defined by the stochastic Langevin equation 
have the same $t \ra 0$ asymptotic expansion which is shown to be 
Borel summable. The Borel transform correctly reproduces the 
time dependent moments of the complex measure for all $t$, 
including their $t \ra \infty$ equilibrium values.  
On the other hand the results of a 
direct numerical simulation of the Langevin moments are found to 
disagree from the `correct' result for $t$ larger than a finite $t_c$. 
The breakdown time $t_c$ increases powerlike for decreasing strength of 
the noise's imaginary part but cannot be excluded to be finite 
for purely real noise. To ascertain the discrepancy we also compute 
the real equilibrium distribution for complex noise explicitly and verify 
that its moments differ from those obtained with the complex measure. 
\end{quote}
\end{center}

\newpage

\newsection{Introduction}

The complex Langevin method \cite{Parisi,Klauder} is arguably the 
best candidate framework to define and compute Lorentzian signature 
functional integrals beyond series expansions. In brief it 
aims at replacing functional averages of some quantity $\cO$ with 
the $e^{i S}$ integrand (`complex measure') by a limit of averages 
computed with a real measure on a `doubled' configuration space,
see \cite{StochQuant} for a review.  
In the case of a one-component scalar field theory 
\be 
\label{i1} 
\dfrac{\int \!\cD \phi \,\cO(\phi) e^{i S(\phi)}}%
{\int \!\cD \phi \,e^{i S(\phi)}}
 = \lim_{\th \ra \pi_-} 
\lim_{t \ra \infty} 
\dfrac{\int\! \cD \phi_r \cD \phi_i \,
\cO(\phi_r + i \phi_i ) R_{t,\th}(\phi_r,\phi_i)}%
{\int\! \cD \phi_r \cD \phi_i \,R_{t,\th}(\phi_r,\phi_i)}\,,
\ee
where $R_{t,\th}$ is a real positive measure for all 
$t$ and $\th$ such that $R_t(\phi_r,\phi_i) \ra 
\delta(\phi_r) \delta(\phi_i)$ for $t \ra 0$. 
The angle $0 \leq \th < \pi$ is related to the 
phase of the Wick rotation and normalized such that $\th=0$ and 
$\pi$ correspond to Euclidean and Lorentzian signature, respectively. 
The right hand side can be evaluated numerically for fixed 
$t$ and $\th$ by 
a two component version of the usual Langevin method.
That is, a pair of stochastic differential equations 
driven by white noise $b_r,b_i$ is solved for each instance
of $b_r,b_i$ and $\cO(\phi^b_r + i \phi^b_i)$ is evaluated on the 
solution $\phi_r^b(t),\phi_i^b(t)$ at $t$, after which the ensemble average 
$\overline{\cO(\phi_r^b + i \phi_i^b)}$ 
yields the ratio in (\ref{i1}) for fixed $t,\th$.

The numerical implementation of the Langevin method is 
seductively simple. The results of 
unguided numerical experiments are however often 
inconclusive: the large time limit of the averages entering the 
right hand side of (\ref{i1}) may fail to exist or may converge 
to the `wrong' answer, in simple cases where the left hand side 
of (\ref{i1}) can be evaluated by other means, see \cite{Aarts1} 
for a recent critical discussion. There is also a considerable 
body of mathematical work on the subject which unfortunately 
does not seem to cover the situations directly relevant 
to (\ref{i1}) and its numerical implementation. What is lacking 
is a theoretical understanding of the domain of validity of the method: 
if it fails, why does it fail, and conversely in what 
circumstances can one be assured that the right hand side 
of (\ref{i1}) indeed evaluates the left hand side. 
 
Although the interest in the complex Langevin method 
comes mostly from field theory gradients are not at the core of the 
issue. In line with earlier 
investigations \cite{Klauder2,Gausterer,Guralnik1,Guralnik2,Bernard} 
we will therefore focus on the zero dimensional case and 
specifically on the paradigmatic 
case of a $\phi^4$ interaction. To appreciate the origin 
of the angle $\th$ and the simplifications made compared  
to a $1\!+\!d$ dimensional $\phi^4$ theory we quickly run 
through the main steps of the Wick rotation in a lattice 
formulation. We start from the discretized Minkowski space 
action on a cylinder $T \times L^d$, with different lattice 
spacings $a_0$, $a$ in the temporal and the spatial 
directions. This gives
\be
\label{i2}
S_{\rm M} = \frac{1}{2}\frac{a^{d-1}}{a_{0}}
\sum_{n}(\Delta_{0}\phi_{n})^{2} - \frac{1}{2}a^{d-3}a_{0}
\sum_{n,i}(\Delta_{i}\phi_{n})^{2}
-\frac{m^{2}}{2}a^{d-1}a_{0}\sum_{n}\phi_{n}^{2}
-\frac{\lambda}{4}a^{d-1}a_{0}\sum_{n}\phi_{n}^{4}\,,
\ee
where $n$ labels the site, $T=N_{0}a_{0}$, $L=Na$, and 
$\Delta_{0}\phi_{n} :=\phi_{n+\hat{0}}-\phi_{n}$, $\Delta_{i}\phi_{n} 
= \phi_{n+\hat{i}}-\phi_{n}$, $i=1,2,..,d$. Replacing 
$a_0 \mapsto - i e^{- i\th/2} a_0$ and reverting to lattice units 
$a_0 = a =1$ the partition function becomes 
\ba
\label{i3}
Z_{\rm \th} &=& \int \prod_{n}d\phi_{n} \exp{(-S_\th)} \,,
\nonum
S_{\rm \th} &:=& 
\frac{1}{2}e^{-i\th/2}\sum_{n}(\Delta_{0}\phi_{n})^{2} 
+ e^{i\th/2}\sum_{n}\Big[\frac{1}{2}(\Delta_{i}\phi_{n})^{2}
+\frac{m^{2}}{2}\phi_{n}^{2}+\frac{\lambda}{4}\phi_{n}^{4}\Big]\,.
\ea
This multidimensional integral over the real variables $\phi_{n}$ 
is absolutely convergent for all $-\pi < \th < \pi$. 
Evaluating the two point function based on (\ref{i3}) for $\lb =0$ 
and $\th = \pi - \eps$ with $\epsilon$ a positive infinitesimal, 
one recovers the discretized Feynman propagator. This suggests that 
suitable distributional limits of the $Z_{\th}$ based correlators
with $\th \ra \pi - \eps$ would in principle define the 
Lorentzian signature lattice theory. 
Setting aside the subtle distributional aspects $Z_{\th}$ is for 
$\th \neq 0$ also not suited for evaluations based on numerical 
stochastic approaches (Monte Carlo, Langevin, etc) as the exponent 
is dominated by the imaginary parts, inducing an intolerably 
low signal to noise ratio. The Euclidean signature version 
corresponds to $\th =0$ and circumvents both awkward features, 
at the expense of a more indirect recovery of the Lorentzian 
signature amplitudes eventually aimed at. The gradients 
in (\ref{i3}) are not central to the problem. By 
discarding them one obtains a zero dimensional system
whose stochastic quantization resembles $1+0$ dimensional 
quantum field theory (i.e.~quantum mechanics) with the stochastic 
time providing the 
added dimension. In the following we write $S(q)$ for a 
(in general complex valued) polynomial action for the real 
variable $q$ which carries a $\th$ dependence induced by (\ref{i3}) 
interpolating between Lorentzian ($\th=\pi$) and Euclidean 
signature ($\th=0$).

Associated with the complex action $S(q)$ is the 
complex Fokker-Planck equation 
\be 
\label{i4} 
\frac{\dd}{\dd t} \rho_t(q) = \bP \,\rho_t \,,\quad 
\bP = \dd_q (\dd_q + \dd_q S) \,,
\ee  
in the stochastic time $t$ with initial condition 
$\rho_{t =0}(q) = \rho_0(q)$. Here $\bP$ is 
the transpose of the usual Langevin operator $\bL = \dd^2_q - \dd_q S \dd_q$ 
and formally $\rho_t(x) = \exp(t \bP) \rho_0$. One assumes that for 
suitable initial date $\rho_t$ exists and reproduces the 
complex Boltzmann factor in the limit $t \ra \infty$, i.e.
$\lim_{t \ra \infty} \rho_t = e^{-S}$. For an `observable'
$\cO(q)$ the average wrt the complex measure $\rho_t$ is 
defined by 
\be 
\label{i5}  
\bra \cO \ket_{\rho_t} = \dfrac{ \int \!dq \, \cO(q) \,\rho_t(q) }%
{ \int \!dx \,\rho_t(q) }\,.
\ee
On the other hand the real measure relevant for the right hand side
of (\ref{i1}) is defined by the following real Fokker-Planck equation 
\ba
\label{i6} 
&& \frac{\dd}{\dd t} R_t(x,y) = \P\, R_t(x,y)\,,
\nonum
&& \P = \dd_x( A_R \dd_x - F_x) + \dd_y( A_R \dd_y - F_y) \,,
\quad A_R - A_I =1\,,
\nonum
&& F_x = - {\rm Re}[\dd_x S(x+iy)]\,, \quad 
F_y = - {\rm Im}[\dd_x S(x+iy)]\,,
\ea
where the forces satisfy $\dd_y F_x + \dd_x F_y=0$ and
$A_R = A_I +1$, $A_I \geq 0$, reflects a variant of the 
fluctuation-dissipation theorem. The initial conditions are 
$R_0(x,y) = \rho_0(x) \delta(y)$, and again one has assume that 
$R_t = \exp(t \P) R_0$ is well 
defined and has a limit for $t \ra \infty$. For analytic 
observables (depending on $x+iy$ only) one considers the 
averages 
\be    
\label{i7} 
\bra \cO \ket_{R_t} = \dfrac{ \int \!dxdy \, \cO(x+iy) \,R_t(x,y) }%
{ \int \!dxdy \,R_t(x,y) }\,.
\ee
The Parisi-Klauder conjecture \cite{Parisi,Klauder} states 
that under `suitable subsidiary conditions' both averages coincide:
\be 
\label{PKC}
\bra \cO \ket_{\rho_t} \stackrel{\displaystyle ?}{=} 
\bra \cO \ket_{R_t} \,,\quad \mbox{for all} \;\;t \geq 0\,. 
\ee
Then $\lim_{t \ra \infty} \bra \cO \ket_{\rho_t} = 
\lim_{t \ra \infty} \bra \cO \ket_{R_t}$ should  
follow and the simulation of the two-component Langevin 
equation can for large $t$ be used to compute the expectation 
values with the complex Boltzmann factor $e^{-S(q)}$ as 
in (\ref{i1}). 

For definiteness we focus on the case of quartic action 
$S(q) = \alpha e^{i \th/2} q^4$, $0< \alpha$, $0 \leq \th <\pi$, 
where the moments $\bra x^p \ket_{\rho_t}$ and $\bra (x+iy)^p \ket_{R_t}$,
with $p \in \N$, fully characterize the underlying measures.   
We show in Section 3 that both sets of moments have identical 
$t \ra 0$ asymptotic expansions of the form
\be 
\label{i8}
\bra x^p\ket_{\rho_t} \sim \sum_{n \geq  p/2\!\! \mod 2} \!\!\!c_{p,n} \; 
(-4 \alpha e^{i\th/2})^{\frac{n-p/2}{2}}\; 
\frac{(2 t)^n}{n!} \sim \bra (x+iy)^p \ket_{R_t}\,,\quad c_{p,n} \in \N\,.
\ee
Moreover the series (\ref{i8}) is Borel summable and defines 
a unique function 
\be 
\label{i9} 
M_p(t) = e^{- i\frac{\th}{4}(1 + \frac{p}{2})}\, t^{-1} 
\int_0^{\infty} \!ds \, \exp\Big(\!-\!\frac{s}{t e^{i\th/4}} \Big)\, b_p(s) \,,
\quad 0 \leq \th < \pi\,,
\ee
where $b_p(s)$ is the Borel sum of (\ref{i8}) for $\alpha >0$. 
Third we show 
\be 
\label{i10} 
M_p(t) = \bra x^p\ket_{\rho_t} \,, \quad \mbox{for all} \;\;t \geq 0\,.
\ee
In other words the Borel resummation of the short time asymptotic 
expansion (\ref{i8}) correctly captures the dynamics of the  
complex measure (\ref{i5}) aimed at, including its equilibrium aspects.

Based on (\ref{i8}) one might expect that the same holds true for the 
real measure defined by (\ref{i6}). However, by direct numerical 
simulation we find 
\ba
\label{i11} 
&& M_p(t) = \bra (x+iy)^p \ket_{R_t} \,,\quad 
0 \leq t \leq t_c(A_I)\,,
\nonum
&& M_p(t) \neq \bra (x+iy)^p \ket_{R_t} \,,\quad 
t > t_c(A_I)\,,
\ea 
where the `breakdown' time $t_c(A_I)$ depends on the strength $A_I$ 
of the imaginary noise. Generically therefore the conjectured 
equality (\ref{PKC}) holds for a finite time interval only, rendering 
its use to define the left hand side of (\ref{i1}) in terms of the 
right hand side problematic. The breakdown time increases powerlike 
as $A_I \ra 0$. Taking $A_I$ strictly zero may lead to an increased
sensitivity on initial conditions in the Langevin simulations 
\cite{Ambjorn,Guralnik1}. A refined version of the conjecture 
(\ref{PKC}) thus has to read 
\be 
\label{PKCA}
\bra \cO \ket_{\rho_t} \stackrel{\displaystyle ?}{=} 
\lim_{A_I \ra 0} \bra \cO \ket_{R_t} \,,\quad 
0 \leq t \leq \lim_{A_I \ra 0} t_c(A_I)\,. 
\ee
For practical purposes (\ref{PKCA}) may suffice provided the 
temporal variations become small before  $\lim_{A_I \ra 0} t_c(A_I)$
is reached \cite{Aarts1}. A theoretical foundation of the method 
however requires a proof that $\lim_{A_I \ra 0} t_c(A_I) = \infty$.

The article is organized as follows. In Section 2 we 
show that the  moments $\bra x^p\ket_{\rho_t}$ admit a 
transfer operator representation in terms of a non-selfadjoint 
propagation kernel $(e^{-t {\bf H}})(q,q')$  whose properties 
we examine in detail. In particular $(e^{-t {\bf H}})(q,q')$  
is shown to admit a well-defined spectral representation 
whose norm-convergence is governed by Davies' spectral norms 
\cite{Daviesbook,Davies1,Davies2}. It follows that the kernel's 
$t \ra \infty$ limit is well-defined and correctly projects onto 
its ground state proportional to $e^{-S(q)/2}$. In Section 3 we 
derive the results (\ref{i8}) 
-- (\ref{i10}) for the short time asymptotics and its Borel 
resummation. Section 4 is devoted to the numerical simulation 
of the moments leading to (\ref{i11}) and the refined conjecture 
(\ref{PKCA}). For $A_I >0$ the numerical moments also have 
finite large $t$ limits which however differ from the correct 
answer $\lim_{t \ra \infty} \bra x^p\ket_{\rho_t}$. To ascertain the 
disagreement we study in Section 5 directly the spectrum and the 
ground state of the real Fokker-Planck operator $\P$. Both are 
seen to be compatible with the working hypothesis that 
$t \mapsto e^{t \P}$ indeed defines a strongly continuous semigroup 
with a pointwise non-negative kernel for all $A_I>0$. Its ground 
state $\vp_0$ is then used to independently compute the asymptotic 
values $\lim_{t \ra \infty} \bra (x+iy)^p \ket_{R_t}$ as 
$\bra (x+iy)^p\ket_{\vp_0}$. Their agreement leaves a non-naive action 
of the semi-group $e^{t \P^T}$ on holomorphic functions as the likely 
culprit for the failure of (\ref{PKC}) for $A_I>0$.   
Two appendices contain supplementary  material: in Appendix A 
the noninteracting case is discussed and the complex Langevin 
method is shown to work perfectly. The under-determination of the 
observable flow mentioned above gives rise to an interesting 
parallelism between the Parisi-Klauder conjecture and 
quantum mechanical `supertasks' \cite{Norton} which we describe in 
Appendix B.

\newpage 
\newsection{The transfer operator for the complex sextic oscillator}

The averages based on the complex measure admit a transfer 
operator realization akin to a quantum mechanical system in 
the stochastic time $t$: 
\be 
\label{rhotrans}
\bra \cO \ket_{\rho_t} = \dfrac{
\int\!dq dq'\, \cO(q) \,e^{S(q)/2} (e^{-t \bH})(q,q') e^{-S(q')/2} \rho_0(q')}%
{\int\!dq dq'\,e^{S(q)/2} (e^{-t \bH})(q,q') e^{-S(q')/2} \rho_0(q')}\,.
\ee
The operators $e^{-t \bH}, \,t>0$, generate a semigroup with integral kernel 
$(e^{-t \bH})(q,q')$ to which we will refer to as the `complex propagation 
kernel'. Since $\bH$ is not selfadjoint for $\th \neq 0$ none of the usual 
properties of a transfer operator semigroup can be taken for granted. 
In Section 2.1 we investigate the spectrum of $\bH$ and in Section 2.2 
the spectral norms governing its norm-convergence properties for 
$t \ra \infty$.   

We begin by defining $\bH$. The complex Fokker-Planck operator 
$\bP$ in (\ref{i4}) is not symmetric even when specialized to a real 
action. By a similarity transformation it can be mapped into 
a conventional Schr\"{o}dinger type operator which is selfadjoint 
for real actions. Starting from $\bP$ in (\ref{i4}) we define
\be 
\label{s0} 
\bH := - e^{S(q)/2} \bP e^{-S(q)/2}\,,\quad \bP e^{-S(q)} =0\,,\quad 
\bH e^{-S(q)/2} =0\,.
\ee
This gives $\bH = p^2 + V_{\rm FP}(q)$ with the ``Fokker-Planck'' potential 
\be 
\label{s2} 
V_{\rm FP} = \frac{1}{4} \bigg( \frac{\dd S}{\dd q} \bigg)^2 - 
\frac{1}{2} \frac{\dd^2 S}{\dd q^2} \,.
\ee 
The associated Schr\"{o}dinger operator factorizes 
$\bH = (-\dd_q + \frac{1}{2}\dd_q S) ( \dd_q +\frac{1}{2}\dd_q S)$  
and has $\exp\{-\frac{1}{2}S(q)\}$ as exact ground state
with zero energy; a feature related to an underlying supersymmetry,
see \cite{Gozzi,StochQuant} for reviews.

For selfadjoint Schr\"{o}dinger operators $\bH = p^2 + V(q)$, with 
$V(q)$ a real even polynomial of degree $2p$, $p \geq 1$, the spectrum 
is known to be positive, purely discrete, and nondegenerate. 
A complete set of real-valued orthonormal eigenfunctions 
$\psi_n, \,n\geq 0$, exists and the transfer operator 
(Euclidean signature propagation kernel) has the spectral 
decomposition 
\be 
\label{s1} 
\big( e^{-\frac{t}{2} \bH }\big)(q,q') = 
\sum_{n \geq 0} e^{-\frac{t}{2} E_n} P_n(q,q')\,.
\ee
The $E_n$ are the eigenvalues of $\bH$ and 
$P_n(q,q) = \psi_n(q) \psi_n(q')$ is the projector 
onto the $n$-th eigenspace. In particular the semigroup 
$t \mapsto \exp\{- t \bH\}$, $t>0$, is strongly continuous 
and the $t \ra \infty$ limit converges strongly to the 
projector $P_0$ onto the ground state. When the couplings 
parameterizing the potential $V$ become complex, the 
operator $\bH$ is no longer selfadjoint and 
none of the above properties can be taken for granted. 
In fact, in addition to the spectrum becoming complex,
several other new phenomena occur for complex couplings,
originally explored by E.B.~Davies in the harmonic case 
\cite{Davies1}.

In the following we consider the family of hamiltonians
\be
\label{s3}
\bH = p^2 - e^{i\frac{\th}{2}} \om^2 q^2 + \lb e^{i\th} q^6\,,\sspace 
0 \leq \om\,,\;\;0< \lb\,,\;\;0\leq \th < \pi\,,
\ee
with $p = -i \dd/\dd q$. The parameterization is chosen such that 
the complex quartic action   
\be 
\label{s4} 
S = \frac{1}{2} \sqrt{\lb} e^{i\th/2} q^4\,,
\ee
gives rise to (\ref{s3}) with $\om^2 =3\sqrt{\lb}$,
and $\th$ is the Wick rotation angle of the original problem. 
We often keep $\om^2$ as an independent coupling 
as many structural properties continue to hold for generic 
$\om^2$. In addition to $\om^2/\sqrt{\lb} =3$ a sequence of 
other coupling ratios, namely $\om^2/\sqrt{\lb} = 4 k+ 2 \nu +1$, 
$k \geq 0$, $\nu = \pm 1$, gives rise to Schr\"odinger equations 
which are `quasi-integrable' in the sense that the first $k$ solutions 
of a given parity $\nu$ have the form $\psi(q) = 
P_{k-1}(q) \exp\{ - \sqrt{\lb} q^4/4\}$, with $P_{k-1}$ a polynomial
of degree $k\!-\!1$. Responsible for this phenomenon is an underlying 
dynamical $sl_2$ symmetry, see \cite{Turbiner} and the references 
therein. We aim at understanding the structure of the spectrum,
of the spectral projections, and of the transfer operator
associated with the hamiltonians (\ref{s3}) as a function of $\th$. 
We begin with the spectrum. 


\newsubsection{Spectrum} 

As mentioned, for $\theta =0$ the spectrum of $\bH$ is positive, 
purely discrete, and non-degenerate.    
For complex couplings the structure of the discrete spectrum 
can be understood from a scaling argument. Rewriting the eigenvalue
equation for $\th=0$, i.e.~$\bH|_{\th =0} \psi = E \psi$, $\om \geq 0$, 
in terms of $z = \exp\{-i \th/8\}q$ one finds 
$\bH \Omega = \exp\{ i\th/4\} E \Omega$, with $\Omega(z) := 
\psi(\exp\{i \th/8\} z)$. Assuming that the angle $\th$ is 
restricted such that normalizability is preserved this 
determines the $\th$-dependence of the eigenvalues. 
Writing $E_n(\om^2,\lb, \th),\,n \geq 0,$ for the eigenvalues 
of $\bH$ the relation 
\be 
\label{s5}
E_n(\om^2,\lb,\th) = e^{i\frac{\th}{4}} E_n(\om^2,\lb,0) \,,\quad \om^2 \geq 0\,,
\;\;\lb >0\,,
\ee 
links the discrete spectra for real and complex couplings. 
Normalizability is preserved for $0 \leq \th < \pi$, as anticipated in 
(\ref{s3}): by substitution into the differential equation one 
sees that normalizable wave functions of a $\lb q^p$ potential have 
a dominant $\exp\{-2\sqrt{\lb} q^{1+p/2}/(p+2)\}$ decay. For $p=6$
and $q = \exp\{i \th/8\} z$ the exponential is damping in $z$ as 
long as $e^{i\th/2} z^4$ has a positive real part. Incidentally 
(\ref{s5}) also provides a means to define the 
discrete spectrum of the `wrong sign' sextic anharmonic 
oscillator as $\lim_{\th \ra \pi_-} E_n(0,\lb,\th) = 
\frac{1}{\sqrt{2}}(1 + i ) E_n(0,\lb,0)$. This leads to 
a discrete spectrum located on the diagonals of the right half 
plane, in contrast to the $E_n(0,\lb,\pi)=E_n(0,-\lb,0)$ which are 
real but unbounded from below.       

To the best of our knowledge the energy levels $E_n(3 \sqrt{\lb}, 
\lb, 0)$ have not been computed before. The  factorization 
$\bH|_{\om = \sqrt{3 \lb}} = (-\dd_q + \sqrt{\lb} q^3)(\dd_q + \sqrt{\lb} q^3)$
shows that 
\be 
\label{s8} 
\Omega_0(q) =  \frac{2^{3/8}}{\Gamma(1/4)^{1/2}} 
\lb^{1/16} \exp\Big\{ - \frac{\sqrt{\lb}}{4} q^4 \Big\}\,,
\ee
is the unique normalized ground state with energy $E_0=0$.   
To get the excited state energies we use a simple but reliable 
technique: for a suitable basis on $L^2$ the matrix elements 
of $\bH$ define a matrix operator. Diagonalizing truncations
of this infinite dimensional matrix produces approximate eigenvalues 
whose accuracy can be tested by probing for truncation 
independence. A natural choice of basis are the Hermite 
functions. For $\lb >0$ they no longer capture the qualitative 
behavior of the exact eigenfunctions but the resulting 
matrices have a band structure with only a few diagonals 
populated which is numerically advantageous.

To compute the corresponding matrix elements we express the 
hamiltonians and the Hermite functions in terms of creation 
and annihilation operators. With the normalizations 
$q = (a^* + a)/\sqrt{2\om}$, $p = i(a^* - a)\sqrt{\om/2}$, $\om>0$, 
$[a,a^*]=1$, the Hermite functions are given by 
$|n \ket = \frac{1}{\sqrt{n!}} {a^*}^n |0\ket$, 
$a |0\ket =0$. Converting (\ref{s3}) into a normal ordered 
expression in terms of $a^*,a$ the matrix elements between 
the hermite states are readily obtained and read
\ba 
\label{s11}
&\nspace & \bra m|\bH |n\ket = \frac{\om}{2}(1 - e^{\th/2}) 
(2 n\! +\!1) \delta_{m,n} - \frac{\om}{2}(1 + e^{\th/2})
[\sqrt{n(n\!-\!1)} \delta_{n-2,m} + \sqrt{m(m\!-\!1)} \delta_{m-2,n}]
\nonum
&\nspace & \quad +  
\frac{\lb}{8\om^3} \sqrt{\frac{m!}{n!}} 
\Big\{ \delta_{m-6,n} + (6 m\! -\!9) \delta_{m-4,n} 
+ 15 (m^2\! -\!m \!+\!1) \delta_{m-2,n} \Big\} + (m\,\leftrightarrow\,n) 
\nonum
&\nspace& \quad + \frac{\lb}{8 \om^3}
\big(20 m^3 + 30 m^2 + 40 m +15\big) \delta_{m,n} \,.
\ea 
The low lying parts of the spectrum can now be computed by 
directly diagonalizing the hamiltonian matrices truncated to 
$0 \leq m,n \leq N$. Stability of the spectrum with increasing 
$N$ indicates the reliability of the approximative result. 

As a test we recomputed the ground state energy of $\bH$ with 
$\om =1$ for $10^{-3} \leq \lb \leq 10^3$, where high accuracy results 
are available in the literature, see e.g.~\cite{Banerjee}. 
For all but very large $\lb$ truncations of $N=150$ are 
sufficient to obtain the eigenvalues to 6 digits accuracy. 
The results are in perfect agreement with those tabulated in 
\cite{Banerjee}. Note that the widely used Hill determinant method 
occasionally fails for sextic potentials \cite{Turbiner}.

In Table 1 we present results for the low lying eigenvalues 
in the `Fokker-Planck' case $\om^2 = 3 \sqrt{\lb}$. Both the 
$\th$-dependence and the $\lb$ dependence can be extracted 
analytically, see (\ref{s5}) and below. The overall structure 
of the spectrum comes out as 
\be
\label{s12}
E_n = C_n \,e^{i\th/4} \lb^{1/4}\,,\quad 
C_n \ra C^6 n^{3/2} \,,\quad 
\lb>0\,,\;\;\;0\leq \th <\pi\,, 
\ee
so that it suffices to know the $C_n$. One has $C_0 =0$ 
and for $n \leq 20$ a $9$ digit accuracy can be achieved with 
truncations $N \leq 500$. We only present the first ten 
to six digits. The limiting behavior as $n \ra \infty$ 
follows from a semiclassical analysis, the constant $C$ is 
known analytically, see (\ref{snorm11}).  

\begin{table}[htb]
\centering
\hspace{-10mm}

\begin{tabular}{|c||c|c|c|c|c|}
\hline
\\[-4.5mm]
$n$  & $1$ & $2$ & $3$ & $4$ & $5$
\\[0.5mm] 
$C_n$ & $1.935\,482$ & $6.298 \,496$ &  $11.680\,971$ 
& $18.042\,635$ & $25.254\,605$  
\\[2mm]
\hline\\[-4.5mm]
$n$  & $6$ & $7$ & $8$ & $9$ & $10$ 
\\[0.5mm] 
$C_n$ & $33.226\,111$ & $41.891\,010$ & $51.197\,908$ 
&  $61.105\,360$   & $71.579\,037$  
\\[2mm]
\hline
\end{tabular}
\caption{\small $C_n$ for the eigenvalues $E_n$ of $\bH$  
with $\om^2 =3\sqrt{\lb}$. The truncation size is $150 \leq N \leq 300$,
the last digits are rounded.} 
\end{table}
%


\newsubsection{The complex propagation kernel}

From the derivation of the phase relation (\ref{s5}) one 
sees that the eigenfunctions $\Omega_n$ of (\ref{s3}) 
are related to those, $\psi_n$, of the selfadjoint $\bH|_{\th=0}$ 
by $\Omega_n(q) = \psi_n( e^{i \th/8} q)$. As a consequence 
the $\Omega_n$ are no longer orthonormal with respect to 
the $L^2$ inner product $\bra \psi, \vp\ket = \int \!dq\, 
\psi(q)^* \vp(q)$. Rather the set $\Omega^*_n, \Omega_n$, 
$n \geq 0$, forms a bi-orthogonal basis in $L^2$ \cite{Daviesbook},
\be
\label{snorm1}  
\bra \Omega^*, \Omega_n\ket = 
\int\!dq\, \Omega_n(q) \Omega_m(q) = \delta_{m,n}\,.
\ee 
The quantities 
\be 
\label{snorm2} 
\bra \Omega_n, \Omega_n\ket = 
 \int \!dq \, \Omega_n(q)^*\Omega_n(q) =: N_n(\th)\,,
\ee
can be interpreted as the norms of projectors 
$P_n \psi= \Omega_n \bra \Omega_n^*,\psi\ket$, which satisfy
\ba 
\label{snorm3}
&& P_n P_m = \delta_{n,m} P_n\,,\quad 
\bra P_n \psi, P_m \psi\ket = \bra \psi,\Omega_n^*\ket 
\bra \Omega_n,\Omega_m\ket \bra \Omega_m^*,\psi\ket\,,
\nonum
&& 1\leq \Vert P_n \Vert := 
\sup_{\psi} \frac{\bra P_n \psi,P_n \psi\ket^{1/2}}%
{\bra \psi,\psi\ket^{1/2}}  = N_n\,.
\ea  
The inequality follows by specializing to $\psi = \Omega_n$, 
the Cauchy-Schwarz inequality gives $\Vert P_n \Vert \leq N_n$,
and specialization to $\psi = \Omega_n^*$ enforces equality. 
The {\it spectral norms} (\ref{snorm2}) originally 
introduced by E.B.~Davies for the complex harmonic oscillator 
encode information about the quasi-spectrum and the norm 
convergence of the heat semigroup generated by the non-selfadjoint 
hamiltonian under consideration. For the complex harmonic 
oscillator it was shown in \cite{Davies1} that 
$\lim_{n \ra \infty} \frac{1}{n} \ln N_n = \gamma(\th) < \infty$,
with an explicitly known constant $\gamma(\th)$. In Appendix A 
we present a simple generating formula for the $N_n$ of the 
complex harmonic oscillator from which the before-mentioned 
asymptotics can also be understood. For anharmonic oscillators 
it is only known that the $N_n$ grow super-polynomially 
\cite{Davies2}. 

Our goal in the following is to determine the rate of growth of 
the $N_n$ for the Fokker-Planck hamiltonian $\bH$ with 
$\om^2 = 3 \sqrt{\lb}$. In a first step we show that for the class 
of hamiltonians (\ref{s3}) the $N_n$ only depend on $\om^2/\sqrt{\lb}$ 
and $\th$,
\be 
\label{snorm4} 
N_n(\om^2,\lb,\th) = N_n\Big(\frac{\om^2}{\sqrt{\lb}}, 1,\th\Big).
\ee
To this end we consider the scaling isometry 
\be 
(S_{\lb} \psi)(q) = \lb^{1/16} \psi(\lb^{1/8} q) \,,\quad 
\bra S_{\lb} \varphi, S_{\lb} \psi\ket = \bra \varphi, \psi \ket\,,
\ee 
and note that $S_{\lb^{-1}} q S_{\lb} = \lb^{-1/8} q$. Hence 
\be 
\label{snorm5} 
S_{\lb^{-1}} \bH S_{\lb} = \lb^{1/4} \,\bH\Big|_{\om^2 \ra \om^2/\sqrt{\lb}, 
\;\lb \ra 1}\,,\quad \lb >0\,.
\ee
Rewriting $\bH \Omega_n = E_n \Omega_n$ as $S_{\lb^{-1}} \bH S_{\lb} 
(S_{\lb^{-1}} \Omega_n) = E_n (S_{\lb^{-1}} \Omega_n)$ and combining 
(\ref{snorm5}) with the fact that the spectrum is nondegenerate
gives 
\be 
\label{snorm6}
E_n(\om^2,\lb) = \lb^{1/4} E_n\Big(\frac{\om^2}{\sqrt{\lb}},1\Big)\,,
\quad  
S_{\lb^{-1}} \Omega_n = \Omega_n\Big|_{\om^2 \ra \om^2/\sqrt{\lb}, 
\;\lb \ra 1}\,.
\ee
On the other hand $S_{\lb}$ is an isometry, so that 
\ba
\label{snorm6a}
&& \bra \Omega_n, \Omega_n\ket = N_n =  \bra \Omega_n, \Omega_n\ket
\Big|_{\om^2 \ra \om^2/\sqrt{\lb}, \;\lb \ra 1}\,,
\nonum
&& \bra \Omega_n^*, \Omega_n\ket = 1 =  \bra \Omega_n^*, \Omega_n\ket
\Big|_{\om^2 \ra \om^2/\sqrt{\lb}, \;\lb \ra 1}\,,
\ea 
which establishes (\ref{snorm4}).

Next we observe that the pointwise defined spectral sum 
\be 
\label{snorm7} 
\big( e^{-\frac{t}{2} \bH }\big)(q,q') := 
\sum_{n \geq 0} e^{-\frac{t}{2} E_n} P_n(q,q')\,,
\ee
with $E_n$ the eigenvalues in (\ref{s5}), is a candidate for 
the kernel of the transfer operator. The $N_n$'s in principle 
then are the coefficients in the expansion of the complex 
partition function 
\be
\label{snorm8}
\Tr[ e^{-\frac{t}{2} \bH} ] 
\stackrel{\displaystyle{?}}{=} 
\sum_{n\geq 0} e^{-\frac{t}{2} E_n} N_n\,.
\ee
Since the `tail' of the partition function corresponding 
to large quantum numbers $n$ should be dominated by 
semiclassical configurations one expects that the 
rate of growth of the $N_n$'s can be extracted from 
a WKB-type evaluation of the partition function.

We begin by reconsidering the eigenvalues.  
For Schr\"{o}dinger operators $p^2 + V(q)$, with $V(q)$ a real 
even polynomial of degree $2p$, $p \geq 1$, the spectrum 
is known to be purely discrete and nondegenerate and to 
scale like $E_n \sim C \,n^{2p/(p+1)} + O(n^{(p-1)/(p+1)})$,
for large quantum numbers $n$, see e.g.~\cite{Berezin}. 
Combined with the scaling law 
(\ref{s5}) one obtains for the eigenvalues $E_n$ of the 
hamiltonians (\ref{s3}) a scaling behavior
\be
\label{snorm10}  
E_n \sim C \, n^{3/2} + O(n^{1/2})\,.
\ee
In the Fokker-Planck case, $\om^2 = 3 \sqrt{\lb}$, it follows from 
(\ref{snorm6}) that the exact eigenvalues have the form 
anticipated in (\ref{s12}), where only the constants $C_n$ 
remain to be determined. One expects their large $n$ behavior to 
be governed by a suitable semi-classical approximation. 
Indeed, application of the SUSY WKB formula \cite{SUSYWKB} gives 
for the semi-classical eigenvalues $\eps_n(\lb)$ the simple 
expression 
\be
\label{snorm11} 
\eps_n(\lb) = C^6 n^{3/2} \lb^{1/4} \,,\quad 
C = \bigg( \frac{\sqrt{\pi} \Gamma(5/3)}{\Gamma(7/6)} \bigg)^{1/4}
\approx 1.14599\,.
\ee
The $\lb$-dependence is evidently of the form mandated by the scaling 
law (\ref{snorm6}) while the $n$-dependence is in accordance with 
(\ref{snorm10}). Comparing with (\ref{s12}) one sees that  
SUSY WKB yields an expression for the limiting constant 
$\lim_{n \ra \infty} n^{-3/2} C_n$. In fact, the $C_n$ approach their 
asymptotic values fairly quickly, rendering (\ref{snorm11}) a good 
approximation to the spectrum. Computing the $C_n$'s as in Table 1
the ratio $C_n/(n^{3/2}C^6)$ comes out as: $0.854\, 484$, $0.999\,311$, 
$0.999\,993$, at $n=1,10,100$, respectively. The standard WKB 
formula, in contrast, gives the correct $n \ra \infty$ asymptotics, 
but a far worse description for small $n$.

Motivated by the good description of the spectrum by the 
SUSY WKB approximation, we also consider the associated wave functions. 
Evaluating the result of \cite{Junker} in the case at hand one finds 
that the quasi-classical eigenfunctions depend for real $\lb$ on $q$ 
only through the combination 
\be 
\label{snorm12}
z = \lb^{1/8} n^{-1/4} C^{-1} q\,,
\ee
such that the domain of definition in $q$ corresponds to 
$z \in [-1,1]$ in $z$. The original norm as defined by a 
$q$-integral with $(n,\lb)$-dependent domain translates into 
\be 
\label{snorm13}
N_n(0) := \frac{2}{\tau_0 C^2} \int_{-1}^1 \!dz\, \Omega_n(z) \Omega_n(z) \,,
\quad \tau_0 = \frac{2 \pi^{1/4} \Gamma[7/6]^{3/2}}{\Gamma[2/3] 
\Gamma[5/6]^{1/2}} \approx 1.84928\,.
\ee
The normalizations are such that $N_n(0) \ra 1$ for 
$n \ra \infty$. However the inner products with $n\neq m$ 
do {\it not} approach zero as $n+m \ra \infty$ with fixed $n-m$. 
The quasi-classical eigenfunctions for real $\lb$ therefore do 
not form an orthonormal set. The explicit expression 
for the $n$-th quasi-classical eigenfunction comes out as 
\ba
\label{snorm14}
&& \Omega_n(z) = \frac{n^{-1/4} \lb^{1/16}}{(1 -z^6)^{1/4} } 
\cos\Big[\frac{\pi}{2} n + n z f(z) + \frac{1}{2} \arcsin z^3\Big]\,,
\nonum
&& f(z) := C^4\, {}_2F_1\Big[-\frac{1}{2}, \frac{1}{6}, \frac{7}{6}; 
z^6 \Big]
= -0.047 \,460 \,z^6 - 0.002\,820 \,z^{12} + O(z^{18})\,.
\ea
Note that $(S_{\lb^{-1}} \Omega_n)(z)$ is $\lb$-independent,
as required by (the quasi-classical counterpart of) (\ref{snorm6}). 
One can now readily restore the $\th$-dependence and 
evaluate (\ref{snorm13}) with integrand $|\Omega_n(z)|^2$,
which defines $N_n(\th)$ in the quasi-classical approximation. 
Straightforward numerical integration then shows convincingly
\be 
\label{snorm15}
\frac{1}{n} \ln N_n(\th) \ra \gamma(\th) < \infty\,.
\ee

As an additional test of (\ref{snorm15}) we also investigate 
the scaling of $\ln N_n$ by direct numerical evaluation using 
an extension of the truncation technique employed for the 
eigenvalues. On account of (\ref{snorm4}) it suffices to 
evaluate the $N_n$'s and their asymptotics for one $\lb$. 
Since the interplay between the different normalizations is 
crucial, we spell out the details here: inserting a resolution 
of the identity in terms of real Hermite functions 
$|k\ket, \, k \in \N_0$, into the eigenvalue equation for 
$\bH$, i.e.
\be 
\label{snorm16} 
\sum_j \bra k| \bH |j\ket \bra j|\Omega_n\ket = 
E_n \bra k|\Omega_n\ket\,,
\ee
one identifies the numerically computed eigenvectors of 
the truncated hamiltonian matrix as 
\be
\label{snorm17} 
v_n^{(j)} = c_n \bra j | \Omega_n \ket\,, \quad 
j =1,\ldots N\,,\;\; n \in \N_0\,,\;\; c_n \in \C\,.
\ee 
Assuming that the exact eigenfunctions $\Omega_n$ form a 
bi-orthogonal basis normalized according to (\ref{snorm1}), 
(\ref{snorm2}) one expects
\be 
\label{snorm18}
\sum_j |v_n^{(j)}|^2 \rra |c_n|^2 N_n\,,\quad 
\sum_j [v_n^{(j)}]^2 \rra c_n^2 \,,\quad N \ra \infty\,.
\ee 
Numerical diagonalization routines typically produce 
eigenvectors normalized to have unit norm in $\C^N$. Based 
on (\ref{snorm18}) the $N_n$ can then be obtained via 
\be 
\label{snorm19}
\Big|\sum_j  [v_n^{(j)}]^2 \Big| \rra \frac{1}{N_n} \,,
\quad N \ra \infty\,.
\ee      
We again consider the Fokker-Planck case $\om^2 = 3 \sqrt{\lb}$ 
in detail. From (\ref{s8}) one computes 
\be 
\label{snorm20} 
N_0 = \frac{1}{(\cos\frac{\th}{2})^{1/4}},\quad
\mbox{ for all}\;\; \lb >0\,. 
\ee
The $N_n, n\leq n_0$, we compute numerically via (\ref{snorm19}). 
In a first step we verify the $\lb$-independence 
for $N=500$. Comparing the results for $\lb =10^{-3}, 10^{-1}, 
1, 10, 10^3$, one finds that whenever the $N_n$'s are 
numerically stable they are also $\lb$-independent to the 
same accuracy. Specifically, for $\th$ less than $\pi/4$ 
at least $N_n, \,n =0, \ldots, 30$, are reliable and 
$\lb$-independent at the $10^{-5}$ level, 
while for larger $\th$ only $N_n, \,n =0, \ldots, 10$, or so 
are reliable and $\lb$-independent at the same accuracy level. 
Next we fix $\lb=1$ and evaluate the $N_n$'s from $N =1000,1300,
1500$ truncations. The results for $\th = \pi/2$ are reported in Table 2.
\bigskip

\begin{table}[htb]
\label{logN}
\centering

\begin{tabular}{|c|cccccccc|}
\hline
$n$ & $0$ & $1$ & $2$ & $3$ & $4$ & $5$ & $6$ & $7$
\\ 
\hline\\[-4.5mm]
$\log N_n$ & 0.08664 & 0.21303 & 0.40745 &0.69222 & 1.02414 &1.3901 &
1.7800 &2.1869 
\\\hline
$n$ & $8$ & $9$ & $10$ & $15$ & $20$ & $25$ & $35$ & $45$  
\\
\hline\\[-4.5mm]
$\ln N_n$ & $2.6063$ & $3.0351$ & $3.4711$ & $5.7206$ & $8.0350$ &  
$10.3840$ & $15.139(1)$  & $20.105(3)$   
\\\hline
\end{tabular}
\caption{\small Logarithm of spectral norms at $\th = \pi/2$.} 
\end{table}

\begin{figure} 
\begin{center}
\includegraphics{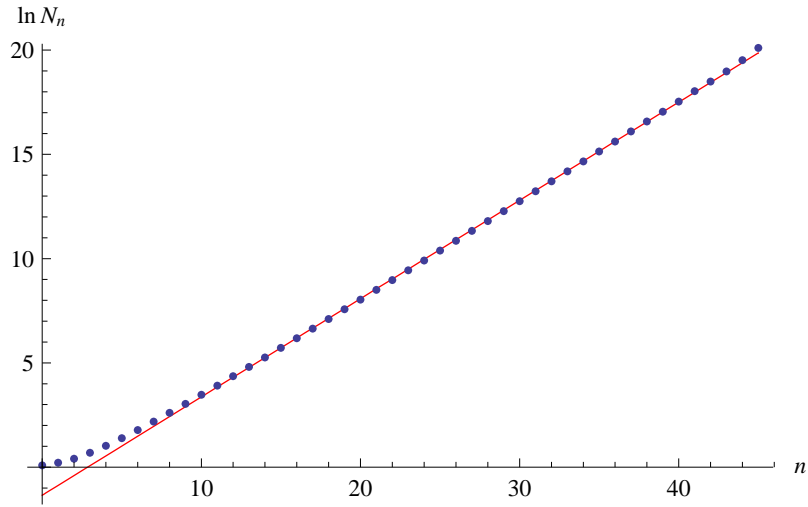}
\end{center}

\caption{\small $\ln N_n$ versus linear fit, $-1.34 + 0.47\, n$.}
\end{figure}

As visible from Fig.~1, a scaling of $\ln N_n$ linear in $n$ is 
favored already for moderately large $n$, consistent with the previous result 
(\ref{snorm15}). 

The scaling law (\ref{snorm15}) has implications for the 
structure of the resolvent and the quasi-spectra of $\bH$. 
In our context (\ref{snorm15}) implies 
that the spectral representation (\ref{snorm7}) is norm convergent
only if $t > t_*$ and divergent otherwise, where using 
(\ref{s12})  
\be
\label{snorm21}
t_* = \frac{2}{\lb^{1/4} \cos\frac{\th}{4}} \max_n \frac{\ln N_n}{C_n}    
<\infty\,.
\ee
Importantly the $t \ra \infty$ limit still projects onto the ground 
state for {\it all} $0 \leq \th <\pi$:  
\be 
\label{snorm22}
s-\lim_{t \ra \infty} e^{-\frac{t}{2} (\bH - E_0) } = P_0\,,
\sspace 
\lim_{t \ra \infty}  e^{-\frac{t}{2} (\bH - E_0)} \psi = 
\Omega_0 \bra \Omega_0^*, \psi \ket\,, \;\;{\rm a.e.}\,.
\ee
This covers the range needed for the Wick rotation and allows one to 
define the Lorentzian signature propagation kernel in terms of 
$\th \ra \pi_-$. The result is somewhat surprising as for 
$\pi/2 < \th <\pi$ the potential in (\ref{s3}) becomes unbounded 
from below.  
In (\ref{snorm22}) the first equation implies the second by 
Cauchy-Schwarz and the strict positivity of $|\Omega_0(q)|$. 
In extension of the results for the complex harmonic oscillator 
\cite{Davies1,Davies2,Boulton} one expects (\ref{snorm15}) 
also to govern the behavior of $(t,\th) \mapsto 
e^{-t \bH}$ as a bounded holomorphic semigroup.

Finally, the analysis leads to an important scaling relation for 
the averages (\ref{rhotrans}). It suffices to consider the monomials 
$\cO(q) = q^p$ with $p$ even. Writing $m_p(t) = \bra q^p\ket_{\rho_t}$ and 
assuming that the initial value distribution $\rho_0(q)$ 
is for complex coupling replaced with $\rho_0(e^{i\th/8} q)$ one finds 
\be 
\label{oscaling2} 
m_p(t) = e^{-ip\th/8} m_p|_{\th =0}(e^{i\th/4} t)\,,
\ee  
by combining (\ref{snorm7}), (\ref{snorm5}) and (\ref{s5}).

\newsection{Borel resummation of the short time asymptotics}

We now resume the investigation of the putative identity
\be 
\label{pkc}
\bra \cO \ket_{\rho_t} \stackrel{\displaystyle ?}{=} 
\bra \cO \ket_{R_t} \,,\quad t \geq 0\,, 
\ee
as surveyed in the introduction. The results of Section 2 
entail that the left hand side of (\ref{pkc}) is well-defined 
for all $t \geq 0$ and converges to the desired $t \ra \infty$ 
average even if the underlying quartic action is complex. 
Our strategy for investigating (\ref{pkc}) is based on the 
fact that both sides have the same $t \ra 0$ asymptotic 
expansion, as detailed in  Section 3.2. For the quartic selfinteraction 
the coefficients of the expansion turn out to be such that the 
series is amenable to a Borel resummation. The Borel transform 
defines a unique function for all $0 \leq \th <\pi$, which is
shown to  coincide with the left hand side of (\ref{pkc}) in 
Section 3.3. Nevertheless this falls short of proving (\ref{pkc}). 
In order to disentangle the issues involved we recap briefly the 
heuristic arguments for the validity of the conjecture.  

\newsubsection{Recap of the Parisi-Klauder conjecture} 

The rationale for the definition of $\P$ in (\ref{i6}) can be 
understood by rewriting it in complex coordinates, 
$z =  x+ iy,\,\bar{z} = x- iy$.
This gives 
\be 
\label{L5}
\P = [\dd_z^2 + \dd_zS \dd_z + \dd_z (\dd_zS)]
+ [\dd_{\bar z}^2 + (\dd_z S)^* \dd_{\bar{z}} 
+ \dd_{\bar{z}} (\dd_z S)^*] + 2(A_R + A_I) \dd_z \dd_{\bar{z}}\,.
\ee
We may assume $S(z)^* = S^*(\bar{z})$, where $S^*$ has complex 
conjugated coefficients, e.g.~$S^*(z) = \sum_{n\geq 0} a_n^* z^n$ for 
$S(z) = \sum_{n\geq 0} a_n z^n$. One sees the form of $\P$ is dictated
largely by the requirement that its acts like $\bP$ on holomorphic 
functions and maps real functions to real functions. The mixing term 
can formally be removed by taking $A_R + A_I =0$, 
i.e.~$A_R = 1/2$, $A_I =-1/2$. This however compromises the 
standard real decomposition of the complex Langevin equation 
and even in the free case leads to equilibrium distributions 
which are not integrable, see Appendix A. In principle the 
mixed term could be replaced with $\nu(z \bar{z}) \dd_z \dd_{\bar{z}}$ 
for any function $\nu$ of one variable without affecting the 
reality of the operator or its action on holomorphic functions.

The standard heuristic argument for the validity of
(\ref{pkc}) proceeds by contour deformation: 
\be 
\label{L6}
\dfrac{\rho_t(x)}{\int\!dx \, \rho_t(x)}  = 
\dfrac{\int \!dy\, R_t(x -iy,y)}{\int\!dx dy\, R_t(x,y)} \,,
\ee
relates both averages provided $z \mapsto R_t(z -iy, y)$ 
is for fixed $y \in \R$ analytic with suitable fall-off:
\ba
\label{L7} 
\int_{\R^2} \! dx dy\, \cO(x + iy) R_t(x,y) \is 
\int_{\R} \! dy \int_{\R + iy} dz\, \cO(z) R_t(z-iy,y) 
\nonum
\is \int_{\R} \! dx \,\cO(x) \int_{\R}dy\,R_t(x-iy,y)\,.
\ea 
In particular, if $R_t(x,y)$ itself has  a $t \ra \infty$ 
limit $\varphi_0(x,y)$, it should obey 
\be
\label{L8} 
\dfrac{e^{-S(x)}}{\int\! dx\, e^{-S(x)}}  = 
\dfrac{\int \!dy\, \varphi_0(x -iy,y)}{\int\!dx dy\, \varphi_0(x,y)} \,. 
\ee
For convenience we note the resulting asymptotic form 
of (\ref{pkc}) explicitly: there exists an integrable 
nonnegative solution $\vp_0(x,y)$ of $\P \vp_0 =0$ such that 
for $\cO$ of polynomial growth 
\be
\label{APKC}     
\bra \cO(x) \ket_{e^{-S}}  \stackrel{\displaystyle ?}{=} 
\bra \cO(z) \ket_{\varphi_0} \,,
\ee
holds, where 
\be
\label{Aavgs} 
\bra \cO(x) \ket_{e^{-S}} := \dfrac{ \int \!dx \, \cO(x) \,
e^{-S(x)}}{\int \!dx \,e^{-S(x)}} \,,\quad
\bra \cO(z) \ket_{\varphi_0} := \dfrac{ \int \!dxdy \, \cO(x+iy) \,
\varphi_0(x,y)}{\int \!dx \,\varphi_0(x,y)} \,. 
\ee
Under the assumption that $R_t(x,y)$ has a spectral resolution of 
the form $R_t(x,y) = \sum_n e^{-t E_n} c_n \varphi_n(x,y)$, 
$\P \varphi_n = E_n \varphi_n$, the large $t$ limit in (\ref{i7}) 
will be dominated by the $n=0$ term and $e^{-t E_0} c_0$ will drop out 
in the ratio (\ref{i7}). The mere existence of a limit 
$\lim_{t \ra \infty} \bra \cO(z) \ket_{R_t} = \bra \cO(z)\ket_{\varphi_0}$ 
therefore only requires ${\rm Re}E_n \geq 0$, for all $n$, 
not necessarily $E_0 =0$. Likewise the heuristic contour shift 
(\ref{L7}) and its limiting version (\ref{L8}) does not require 
$E_0 =0$. On general grounds $E_0 =0$ must lie in the 
spectrum of $\P$, see (\ref{Pspec2}) below, and $E_0=0$ is also 
the only ground state energy of $\P$ 
compatible with (\ref{APKC}). In contrast to $\bP$ in (\ref{i4}) 
where $e^{-S}$ manifestly is a zero mode, the solutions of 
$\P \varphi_0 = E_0 \varphi_0$
with $\varphi_0$ integrable and non-negative can in general not 
be found analytically, for any candidate $E_0$, so the mere 
existence of an appropriate $\vp_0$ in the kernel of $\P$ is 
nontrivial.

The putative ground state wave function 
$\varphi_0$ has to obey an additional consistency condition. 
Denoting by $\P^T$ the real adjoint of $\P$ one has 
\ba 
\label{L13}
&& E_0 \int\!dx dy \, \cO(x+iy) \, \varphi_0(x,y) = 
\int\!dx dy \, (\P^T \cO)(x+iy) \, \varphi_0(x,y)\,,
\nonum
&& \sspace (\P^T \cO)(z) = 
\dd_z^2 \cO - \dd_z S \, \dd_z \cO\,,
\ea
using only $\P \varphi_0 = E_0 \varphi_0$ and (\ref{L5})  That is: for 
$E_0 \neq 0$ the $\bra \;\;\ket_{\varphi_0}$ averages of 
a holomorphic observable $\cO$ and its `dual' 
$E_0^{-1}\, \P^T \cO$ have to coincide. For $E_0 =0$ 
the $\bra \;\;\ket_{\varphi_0}$ averages of all observables 
in the image of $\P^T$ have to vanish; see \cite{Aarts1} 
for an alternative derivation.  

In fact only the $E_0 =0$ version of (\ref{L13}) 
is compatible with the validity of (\ref{APKC}). To see this,
take $\cO(z) = z^2$. Then (\ref{L13}) reads 
$(E_0/2) \int\! dx dy\, (x\!+\!iy)^2 \,\vp_0(x,y) = 
\int\! dx dy\,\vp_0(x,y) - 
 \int\! dx dy\, (z \dd_z S)(z\!=\!x\!+\!iy) \,\vp_0(x,y)$, 
where by assumption the $\bra \;\;\ket_{\vp_0}$ averages 
can be replaced with $\bra \;\;\ket_{e^{-S}}$ averages. 
This gives 
\be 
\label{L14} 
\frac{1}{2} E_0 \bra x^2\ket_{e^{-S}} = 1 - \bra x \dd_x S \ket_{e^{-S}}\,.
\ee
Upon integrations-by-parts the right hand side vanishes, enforcing 
$E_0=0$ as the only candidate ground state energy for $\P$ compatible 
with (\ref{APKC}). From a general functional analytical principle 
one can in fact obtain $E_0=0$ irrespective of the validity of 
(\ref{APKC}), see (\ref{Pspec2}).

Returning to (\ref{pkc}) and the heuristic argument (\ref{L7})
for it, we stress that even in very simple interacting theories one 
has no analytic control over $R_t(x,y)$, not even for real
arguments and the analyticity assumption is little more than a 
leap of faith. In order to highlight the nontrivial nature of the 
seemingly innocuous steps in (\ref{L7}) we spell out 
some of the mathematical underpinnings needed for
$R_t(x,y)$ to be well-defined. The defining 
relation in (\ref{i6}) is a generalized heat equation
and the operator $\P$ should generate the associated
semigroup. As such it must satisfy a number of necessary 
conditions which we first list and then comment on:   
\begin{itemize} 
\item[(i)] $\P^T \cO(x+iy) = [(\dd_z^2 - \dd_z S \dd_z) \cO](x+iy)$.
\item[(ii)] $\P$ generates a strongly continuous semi-group 
$t \mapsto e^{t \P}:L^1 \ra L^1$, whose kernel $(e^{t \P})(x,y;x',y')$ 
is pointwise positive. 
\item[(iii)] $\P$ has a unique positive ground state 
$\varphi_0 \in L^1$ with zero energy, $\P \varphi_0 =0$.  
\end{itemize} 
Condition (i) is necessary for the differentiated version 
of (\ref{pkc}) to hold, see Section 3.1. It is manifestly 
satisfied by the proposed real operator $\P$ in (\ref{i6}),
but does not uniquely determine it. 
Since $\P$ is not symmetric the natural functional 
analytical setting is that of a dual pair of Banach spaces,
where $\P$ acts on one space and $\P^T$ on its dual. 
Interpreting $e^{t \P}$ as an operator on a weighted 
$L^1$ space a sufficient condition for boundedness is
\be 
\sup_{x',y'} w(x',y') 
\int\!dx dy\,w(x,y)^{-1} (e^{t \P})(x,y;x',y') <\infty\,,
\ee
for {\it real} $x,y,x',y'$ and a suitable weight function 
$w: \R^2 \ra \R_+$ to be specified later. The initial conditions 
$R_0(x,y) = \rho_0(x) \delta(y)$ do not lie in $L^1$ but we
assume that $R_t(x,y)$ is in $L^1$ and is smooth for all $t>0$. 
The requirement that $\P$ is the generator of a strongly
continuous semigroup poses strong functional analytical 
constraints, which are however indirectly coded in the resolvent 
and not verifiable by inspection of the differential operator. 
The same holds for the even stronger condition that 
the evolution kernel is pointwise nonnegative. For $R_t(x,y)$ both 
properties in principle follow from the stochastic differential 
equation provided a global solution with the appropriate initial 
conditions exist. A necessary condition for (ii) is that 
$\P$ has a spectrum of the form 
\be
\label{Pspec} 
\sigma(-\P) = \{ E_r \pm i E_i\,,\; E_r,\,E_i \geq 0\}\,,
\quad E_0 := \inf E_r \in \sigma(-\P)\,.
\ee  
Generally, if $\P^T$ is the Banach space adjoint of $\P$
with respect to a suitable pairing the full spectra of $\P$ 
and $\P^T$ in principle coincide, see e.g.~\cite{ReedSimon}, 
Thm.~VI.7. An eigenvalue $E_n$ of $\P$ however can either be an eigenvalue
of $\P^T$ or lie in $\P^T$'s residual spectrum. Here we interpret 
the operators $\bP^T$ and $\P^T$ as maps from $L^{\infty} 
\ra L^{\infty}$, where $L^{\infty}$ is dual to a weighted $L^1$ 
space. For the dual Banach spaces we take  
\be 
\label{Pspec3} 
L^1(\R^2, w^{-1} dx dy)\,, \quad 
L^{\infty}(\R^2, w dx dy)\,,\quad w(x,y) = e^{-\delta(x^2+y^2)}\,,
\;\;\delta >0\,,
\ee
and similarly for the one variable case. 
The choice of the weight function ensures that all 
polynomials $x^p,y^p, (x+iy)^p$ and also $e^{j_1 x+j_2 y}$,
$e^{j(x+iy)}$ for real sources are elements of the dual 
space. The integrability condition on the eigenfunctions 
of $e^{t \P}$ and $R_t(x,y)$ is stronger than the unweighted 
one, but should for polynomial actions be easily satisfied. 
In particular the constants are elements of 
the dual space and are annihilated by $\bP^T$ and 
$\P^T$, respectively. Hence $0 \in \sigma(\P^T)$ which 
explains why zero lies in the spectrum of $\P$. 
More generally one has the following relations among the 
spectra
\be
\label{Pspec2}
0 \in \sigma(\bH) = \sigma(-\bP) = \sigma(-\bP^T)\,,\quad  
0 \in \sigma(-\P^T) = \sigma(-\P)\,, 
\ee
but $\sigma(-\bP^T) \subset \sigma(-\P^T)$ does not follow. 
Note that all spectra refer to different Hilbert or Banach 
spaces.

Finally, condition (iii) means that the ground state 
energy $E_0$ lies in the discrete spectrum of the operator.
This could be violated in principle but in the present 
setting of scalar fields with a polynomial interaction 
it is a feature one would expect to hold. For a symmetric 
semi-group uniqueness of a normalizable ground state 
follows from strict positivity of the kernel in (ii). 
For a non-symmetric semigroup it hinges on the 
validity of the $t \ra \infty$ projection property. 
Since this will be relevant later on let us assume 
that the kernel in (ii) admits a spectral representation 
of the form 
\be 
\label{Pspec4}  
(e^{t\P})(x,y;x',y') = \sum_{n\geq 0} e^{t E_n} \vp_n(x,y) \psi_n(x',y') 
\ee
where $\vp_n \in L^1$, $\psi_n \in L^{\infty}$, 
$\int\! dx dy \, \vp_m(x,y) \psi_m(x,y) = \delta_{m,n}$ for 
$m,n \neq0$, and $E_0=0$. For the kernel of $e^{t \P^T}$ then the same 
expansion holds with the roles of $\vp_n$ and $\psi_n$ interchanged. 
The relevant projection property reads  
\be 
\label{Pspec5} 
R_t(x,y) = \sum_n e^{t E_n} \vp_n(x,y) \int\!dx' \rho_0(x') \psi_n(x',0) 
\;\ra\; \vp_0(x,y)  \int\!dx' \rho_0(x') \psi_0(x',0) \,. 
\ee

\newsubsection{Observable flow and its small $t$ expansion} 

Under the assumption that $R_t(x,y)$ has sufficient fall-off 
in $x,y$ one can integrate 
by parts and have the transpose (real adjoint) of $\P$ act 
on the observables. By definition of $\P$ its transpose acts 
on holomorphic observables like ${\bf L}_z = \dd_z^2  - \dd_z S \dd_z$. 
This gives rise to a flow equation for the averages 
$\bra \cO\ket_{R_t}$. Indeed, 
\ba
\label{oflow1}
&& \dd_t \int\!dxdy\, \cO(x+iy) \, R_t(x,y) = 
\int\!dxdy\, \cO(x+iy) \, \P R_t(x,y) 
\nonum
&& 
= \int\!dxdy\, \P^T \cO(x+iy) \,R_t(x,y) = 
\int\!dxdy\, ({\bf L}_z \cO)(x+iy) \,R_t(x,y) \,.
\ea
In particular the normalization factor $\int\!dxdy\, R_t(x,y)$ 
is time independent. The same derivation works for 
averages $\bra \cO\ket_{\rho_t}$ starting from the 
complex Fokker-Planck equation $\dd_t \rho_t = \bP \rho_t$. 
Both averages therefore obey itentical systems of 
`observable flow equations': 
\ba
\label{oflow2} 
\dd_t \bra \cO\ket_{R_t} = \bra \dd_z^2 \cO\ket_{R_t} - 
\bra \dd_z S \dd_z \cO\ket_{R_t} \,.
\nonum
\dd_t \bra \cO\ket_{\rho_t} = \bra \dd_x^2 \cO\ket_{\rho_t} - 
\bra \dd_x S \dd_x \cO\ket_{\rho_t} \,.
\ea
The initial data coincide by assumption $\bra \cO\ket_{R_0} = 
\bra \cO\ket_{\rho_0}$ yielding the following simple but fruitful 
Lemma. 
\medskip

{\bf Lemma:} The averages $\bra\cO\ket_{R_t}$ and $\bra\cO\ket_{\rho_t}$ admit 
asymptotic expansions for $t \ra 0$ which coincide 
\be 
\label{lemma}
\bra \cO \ket_{R_{t}} \sim \sum_{n \geq 0} \frac{t^n}{n!} 
c_n \sim \bra \cO \ket_{\rho_{t}} \,,\quad c_n = \frac{\int \!dx \, (\bL^n \cO)(x) 
\rho_0(x)}{\int\! dx\, \rho_0(x)} \,,
\ee
where $\bL = \dd_x^2 - \dd_x S \dd_x$. Generally,  to the extent 
the flow equations (\ref{oflow2}) determine the averages they 
must coincide. 
\medskip

We add some remarks:
(i) The `observable flow equations' generalize 
the Schwinger-Dyson equation for the partition function 
in \cite{Guralnik1,Guralnik2}. Only the equilibrium aspects 
(vanishing time derivatives in (\ref{oflow2})) are 
utilized in the approach of  \cite{Guralnik1,Guralnik2}.
(ii) Clearly both averages coincide as formal power series in $t$. 
In the formal expansion of the $R_t$ averages $\P^T$ acts like 
$\bL_z = \dd_z^2 - \dd_z S \dd_z$ on $\cO(z)$ while in the 
formal expansion of the $\rho_t$ averages $\bP^T$ acts like 
$\bL$, both evidently producing the same coefficients. The 
$c_n$ are also the unique solution of the recursion relations
entailed by inserting a power series ansatz into (\ref{oflow2}).  
By the smoothness property of $e^{t \P^T}$ and $e^{t \bP^T}$ images 
for $t>0$ the exact averages do admit a series expansion in $t$,
and its asymptotic nature readily follows from the uniqueness of the 
$c_n$.   
(iii) In the noninteracting case detailed in Appendix A 
the flow equations (\ref{oflow2}) completely determine the averages,
implying the validity of the Parisi-Klauder conjecture 
without the need to explicitly compute the respective 
propagation kernels. 
(iv) In interacting situations the flow equation (\ref{oflow2}) will in 
general fall short of fully determining the flow of the averages
of generic $\cO$'s.  Rather upon choosing a `suitable complete set' 
of observables the averages of a small subset will recursively determine 
all others. The quartic case detailed later on is a good illustration.
It does not help that for both averages the time dependence can 
formally be attributed to the observables \cite{Aarts1} 
$\dd_t \cO(t) = (\dd_z^2 - \dd_z S \dd_z)\cO(t)$ differing only in the 
interpretation as functions of $z$ and $x$; the under-determination 
persists. 
(v) A two step approach to lift the under-determination is described 
in \cite{Guralnik1,Guralnik2}. In a first step one characterizes solutions 
of the equilibrium Schwinger-Dyson equations as complexified path 
integrals and then tries to select the one corresponding to the 
complex Langevin  process driven by $\P$ via minimization of an 
effective potential. The approach does not discriminate between 
different values of $A_I>0$. 
(vi) Our strategy focusses on situations where the series (\ref{oflow7}) 
is Borel summable, i.e.~when the $c_n$ are sign-alternating and 
of essentially factorial growth.  Then both averages can differ 
only in a fairly prescribed way and the Borel transform of 
the series potentially coincides with the exact result for one 
or both of the averages.

In the framework of the observable flow the equality (\ref{PKC}) 
comes about as follows:
\ba
\label{oflow4}
&& \int\!dxdy\, \cO(x+iy) \, R_t(x,y) = 
\int\!dxdy\, \cO(x+iy) \, [e^{t\,\P} \rho_0(x) \delta(y)] =
\nonum
&& = 
\int\!dxdy\, [e^{t \,\P^T} \cO(x+iy)] \, \rho_0(x) \delta(y) 
\stackrel{\displaystyle{?}}{=}
\int\!dx\, [e^{t \,\bL_x} \cO(x)] \, \rho_0(x) 
\nonum
&&  = \int\!dx\, \cO(x) \,  [e^{t \,\bP} \rho_0(x)] 
= \int\!dx\, \cO(x) \,  \rho_t(x)\,. 
\ea 
In contrast to the traditional argument (\ref{L7}) the 
above variant does not require control over the the kernel 
$R_t(x,y)$ in the complex plane. The only questionable 
step now is the one marked with $?$. 

The transpositions entering the other steps should be 
unproblematic. For the one variable semigroup this 
is because $\bP$ is similar to the real symmetric operator 
$\bH$. One has $- e^{S/2} \bP e^{-S/2} = \bH = \bH^T 
= - e^{-S/2} \bP^T e^{S/2}$, with $\bP^T = \bL_x$, which implies 
for the kernels $(e^{-t \bH})(x,x') = (e^{-t \bH})(x',x)$ and 
\be 
\label{oflow6} 
(e^{t \bP})(x,x') = e^{S(x)/2} (e^{-t\bH})(x,x') e^{-S(x')/2} = 
e^{S(x')} (e^{t \bP})(x',x) e^{-S(x)} = 
e^{t \bP^T}(x',x)\,.
\ee
For the two variable case the kernel of the transposed operator 
is simply $(e^{t \P^T})(x,x';y,y')= (e^{t \P})(y,y';x,x')$,
and granting the correct domains the second step in (\ref{oflow7}) 
follows. In contrast to $\bP$ the kernels of $e^{t \P}$ and 
$e^{t \P^T}$ (with the same order of arguments) can however 
not be related by a similarity transformation with a function 
(multiplication operator). Assuming otherwise and inserting an 
appropriate ansatz one finds $A_I\dd_x F_y = A_R \dd_y F_x$ as 
a necessary condition. Since $\dd_x F_y = - \dd_y F_x$ follows from the 
defining relations, this could hold only for $A_R + A_I =0$. 
Similarity transformations with other operators 
are not excluded. 

The step $?$ holds in the sense of an asymptotic series in $t$ on account 
of the Lemma. For the action of the semigroups themselves, with the image 
supposed to be smooth functions in $t$, we separately 
highlight the corresponding property: 
\begin{itemize} 
\item[(iv)] $\int \!dx' dy' \,\exp(t \P^T)(x,0;x',y') \cO(x'+iy') 
= \int\! dx' \,\exp(t \bP^T)(x,x') \cO(x')$, 
\\[2mm]
for all $t >0$ and all $x$. 
\end{itemize} 
We regard the invalidity of this extension as the 
likely culprit for the failure of the method whenever it fails.
The right hand side is an $A_I$ independent function of $x$ 
for all $t$ while the kernel on the left hand side is 
$A_I$ dependent. In the $t \ra \infty$ limit 
the $x$ dependence disappears but a mismatched normalization 
may have been picked up. The expected $t \ra \infty$ limit of
(iv) can be inferred from (\ref{oflow6}) and (\ref{Pspec5}) 
\be
\label{oflow7} 
\psi_0(x,0) \int\!dx'dy'\, \vp_0(x',y') \cO(x'+iy') = 
\bra \cO \ket_{e^{-S}}\,.
\ee
Consistency with (\ref{APKC}) requires that $\psi_0(x,0) =\psi_0>0$ is 
a constant given 
by $\psi_0^{-1} = \int\!dx'dy' \, \vp_0(x',y')$. Only if both $\P$ and 
$\P^T$ have purely discrete spectra does the uniqueness of $\vp_0$ 
imply the uniqueness of $\psi_0$ and since a constant trivially 
lies in the kernel of $\P^T$ constancy in $y$ follows as well: 
$\psi_0(x,y) = \psi_0$. 

In Section 5 we shall construct $\vp_0$ explicitly for the 
quartic selfinteraction and find that it depends nontrivially 
on $A_I$ and so do its averages of holomorphic observables. 
Our failure diagnostics differs from the one in 
\cite{Aarts1}, Section 4.2,  where the rapid growth of   
$\int \!dx' dy' \,\exp(t \P^T)(x,y;x',y') \cO(x'+iy')$ 
in $y$ (for the $U(1)$ link model) is argued to invalidate 
the integration by parts in the second equality of (\ref{oflow4}).
We essentially define $e^{t\P}$ by $e^{t \P^T}$ acting on 
holomorphic observables so that validity of this step is 
built in. Table 5 in Section 5 also provides some direct 
computational evidence for its legitimacy. Nevertheless 
(\ref{oflow7}) fails because the ground state $\vp_0$ has 
the wrong structure. Thus (iv) must fail irrespective 
of any growth property in $y$ or invalid integration by parts.


\newsubsection{Temporal Borel resummation for quartic actions}

For a polynomial action like $S(x) = \alpha x^4$, 
$\alpha = \sqrt{\lb} e^{i\th/2} /2$, the powers $x^p$, $p=0\! \mod 2$, 
are a natural complete set of observables. We set 
\be 
\label{borel1}
m_p(t) = \left\{ \begin{array}{cl} 
\bra z^p \ket_{R_t} & \quad \mbox{real Fokker-Planck evolution} \,, 
\\[2mm]
\bra x^p \ket_{\rho_t} & \quad \mbox{complex Fokker-Planck evolution} 
\end{array} \right.\,.
\ee
The flow equations (\ref{oflow2}) translate into
$\dd_t m_p = p(p-1) m_{p-2} - 4 \alpha p m_{p+2}$, $p = 0 \!\mod 2$. 
Since $m_0(t) =1$, one sees that all $m_p(t)$, $p \geq 4$, are determined 
by $m_2(t)$ via the recursion   
\be 
\label{borel2} 
m_p = \frac{1}{4 \alpha} \Big[ (p-3) m_{p-4} - 
\frac{1}{p-2} \dd_t m_{p-2} \Big]\,.
\ee
The first few read 
\ba
\label{borel3}
m_4(t) \is \frac{1}{8 \alpha} [ 2 - \dd_t m_2] \,,
\nonum
m_6(t) \is \frac{1}{128 \alpha^2} [ \dd_t^2 m_2 + 96 m_2 \alpha]\,,
\nonum
m_8(t) \is -\frac{1}{3072 \alpha^3} [\dd_t^3 m_2 +576 \alpha \dd_t^2 m_2
-960 \alpha]\,. 
\ea 
The flow equations (\ref{oflow2}) therefore entail that all 
moments in (\ref{borel1}) coincide for all $t$ if the second 
moments coincide, $\bra z^2 \ket_{R_t}= \bra x^2 \ket_{\rho_t}$,
for all $t$. This structure gives rise to an amusing parallelism 
to the quantum mechanical supertasks introduced by 
Norton \cite{Norton} which we outline in Appendix B.

Either by solving (\ref{borel2}) or directly from (\ref{oflow7}) 
one can work out the formal series to any desired order. 
One finds the structure 
\be 
\label{borel4}
m_p(t) = \sum_{n \geq  p/2\! \mod 2} \!\!\!c_{p,n} \; 
(-4 \alpha)^{\frac{n-p/2}{2}}\; 
\frac{(2 t)^n}{n!}\,,\quad c_{p,n} \in \N\,.
\ee
Inserted into (\ref{borel2}) this gives the recursion relations 
\ba 
\label{borel4a} 
c_{p,p/2} \is \frac{p(p-1)}{2} c_{p-2,p/2-1} \,,
\nonum
c_{p,n} \is - (p-3) c_{p-4,n} + \frac{2}{p-2} c_{p-2,n+1}\,,
\quad n > p/2\,.
\ea
In particular $c_{p.p/2} = 2^{-p/2} p!$, so that 
$m_p(t) = p! \,t^{p/2}/(p/2)! + O(t^{p/2 +1})$. 
The coefficients relevant for the second and fourth moment
are related by $c_{4,n} = c_{2,n+1}$, and the first few are listed 
in Table 3, later on the first nonzero $500$ are used. 

\begin{table}[htb]
\centering

\begin{tabular}{|c|l|}
\hline
$c_{2,1}$ & $1$ 
\\
$c_{2,3}=c_{4,2}$ & $6$ 
\\
$c_{2,5}=c_{4,4}$ & $216$ 
\\
$c_{2,7} = c_{4,6}$ & $22896$ 
\\
$c_{2,9} = c_{4,8}$ & $5360256$
\\
$c_{2,11} = c_{4,10}$ & $2346299136$
\\\hline
\end{tabular}
\caption{\small Coefficients in the asymptotic 
series (\ref{borel4}) for $p=2,4$.} 
\end{table}

According to the Lemma the series (\ref{borel4}) should 
be asymptotic to the exact result for $t \ra 0$. 
By direct summation of partial sums up to $O(t^N)$ 
one finds that the result is $N$ independent for small 
$0\leq t < t_*(p)$, where $t_*(p) \approx 0.13$ for $p=2,4$. 
For small $t$ the series thus defines a function whose 
values can be compared with the numerical simulations. 
One finds an excellent agreement  
\be 
\label{borel5}
m_p(t) |_{\rm series} = m_p(t)|_{\rm simulation}\,,\quad 
0\leq t < t_*(p)\,,\quad p=2,4\,. 
\ee
This corroborates the Lemma and more specifically determines an 
interval $0\leq t < t_*(p)$ in which the series (\ref{borel4}) 
provides a valid description of the exact functions 
$m_p(t)$ in (\ref{borel1}). Assuming that $\inf_p t_*(p) >0$,
this also leads to a strategy to prove the Parisi-Klauder 
conjecture for short times based on the series (\ref{borel4}). 

Next consider the growth rate of the coefficients. With the 
parameterization
\be
\label{borel6} 
\log \frac{c_{p,n}}{n!} = \alpha_p (k+1/2) \ln k - \beta_p k\,,
\quad n= p/2 -2 + 2 k,\;\; k \in \N\,,
\ee
one finds $\alpha_p,\beta_p$ close to unity for $p=2,4$ based 
on the $c_{p,n}$ with $p/2 \leq n \leq 1000$. A proof for all 
$n$ can be based on the recursion relations (\ref{borel4a}). 
Since $\ln n! \sim (n+1/2)\ln n -n$ this in combination with the 
alternating signs
indicates that Borel resummation techniques are applicable. 
In choosing the parameterization (\ref{borel6}) we attributed 
powerlike terms to a redefinition of $t$ via 
$t \mapsto (4 \sqrt{\alpha} t)^2$.    

With this understanding the conventional Borel sum rather than the 
Borel-Leroy generalization is applicable. We therefore define 
\ba 
\label{borel7}
b_p(s) &:=& \sum_{n \geq  p/2\! \mod 2}\!\!\! c_{p,n} \; 
(-4 \alpha)^{\frac{n-p/2}{2}}\; 
\frac{(2 s)^n}{n!^2}\,,
\nonum
M_p(t) &:=& t^{-1} \int_0^{\infty} \!ds \, e^{-s/t}\, b_p(s) \,,
\ea
initially for real $\alpha >0$.  
One then finds that the partial sums $0 \leq n \leq 1000$ produce
truncation independent results for the Borel sums $b_2,b_4$ 
in the interval $0 \leq s \leq 11$, as shown in the inserts of 
Figures 2,3. The restricted integration produces a 
well-defined Borel transform in the interval $0 \leq  t \leq 2$, 
shown in Figures 2,3 (solid lines). The Borel transforms extend the  
low $t$ regime of the original functions in (\ref{borel5}). 
and quickly  approach constant values at around $t=1$. 
Importantly they also agree with the averages defined 
by the complex propagation kernel: 
\be 
\label{borel8} 
M_p(t) = \bra x^p \ket_{\rho_t}\,\quad \mbox{for all} \;\;t \geq 0\,.
\ee
For $\alpha>0$ the equality follows from the uniqueness of 
the Borel transform subject to (\ref{borel6}).    
The extension to complex couplings can be 
done via scaling relations. Generally (\ref{borel7}) implies 
\be
\label{borel9}
M_p(t) = \alpha^{-p/4} M_p\big|_{\alpha=1}(\alpha^{1/2} t)\,,
\ee
provided $M_p(t)$ is analytic in $0 \leq {\rm arg}\,t < \pi/4$. 
For the Borel sum analyticity in $0 \leq {\rm arg}\,s < \pi/4$
is manifest and allows one to rewrite (\ref{borel9}) in a 
way such that only the Borel sum for real arguments enters
\be 
\label{borel10} 
M_p(t) = e^{- i\frac{\th}{4}(1 + \frac{p}{2})}\, t^{-1} 
\int_0^{\infty} \!ds \, \exp\Big(\!-\!\frac{s}{t e^{i\th/4}} \Big)\, b_p(s) \,,
\ee
where $b_p(s)$ is the Borel sum for $\alpha >0$ and $M_p(t)$ 
represents the moments for $e^{i\th/2} \alpha$. On the other hand 
(\ref{borel9}) matches precisely the scaling relation 
(\ref{oscaling2}) derived in Section 2 from the properties
of the complex propagation kernel. This shows (\ref{borel8}) for 
all $0 \leq \th <\pi$.

In contrast to the setting in Section 2 the Borel transform 
(\ref{borel10}) allows for a direct computational implementation. 
The relation (\ref{borel10}) therefore provides an {\it alternative} 
to the complex Langevin method in the case at hand. It directly 
gives the {\it time dependent} moments for (\ref{rhotrans}) which 
we will compare in Section 4 with the results obtained from the 
complex Langevin simulations.

A compelling numerical demonstration of (\ref{borel8}) for 
$\alpha>0$ is obtained by comparing the results of a 1d Langevin 
simulation for $\bra x^p \ket_{\rho_t}$, $\alpha >0$ fixed, with the 
corresponding Borel transforms. The results are shown 
for $p=2,4$ in Figures 2,3. One also sees that the asymptotic 
values are approched quickly
$\lim_{t \ra \infty} \bra x^2 \ket_{\rho_t} = \bra x^2 \ket_{e^{-S}} 
= 0.47798$ and $\lim_{t \ra \infty} \bra x^4 \ket_{\rho_t} = 
\bra x^4 \ket_{e^{-S}} = 0.5$, respectively.

\begin{figure}[htb]
\begin{center} 
\includegraphics{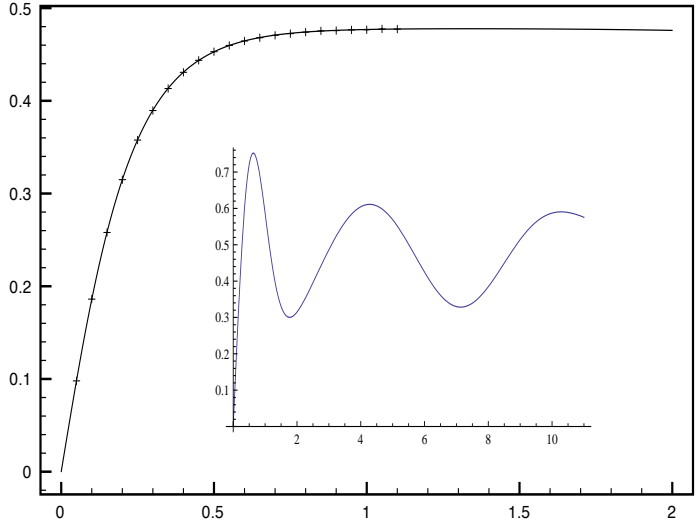}
\end{center} 

\caption{\small Results from 1d Langevin simulation at $\th=0,\lb =1$ 
for $m_2(t)$ versus the Borel transform $M_2(t)$. The insert
shows the Borel sum $b_2(s)$.} 
\end{figure}

\begin{figure}[thb]
\vspace{1cm} 

\begin{center}
\includegraphics{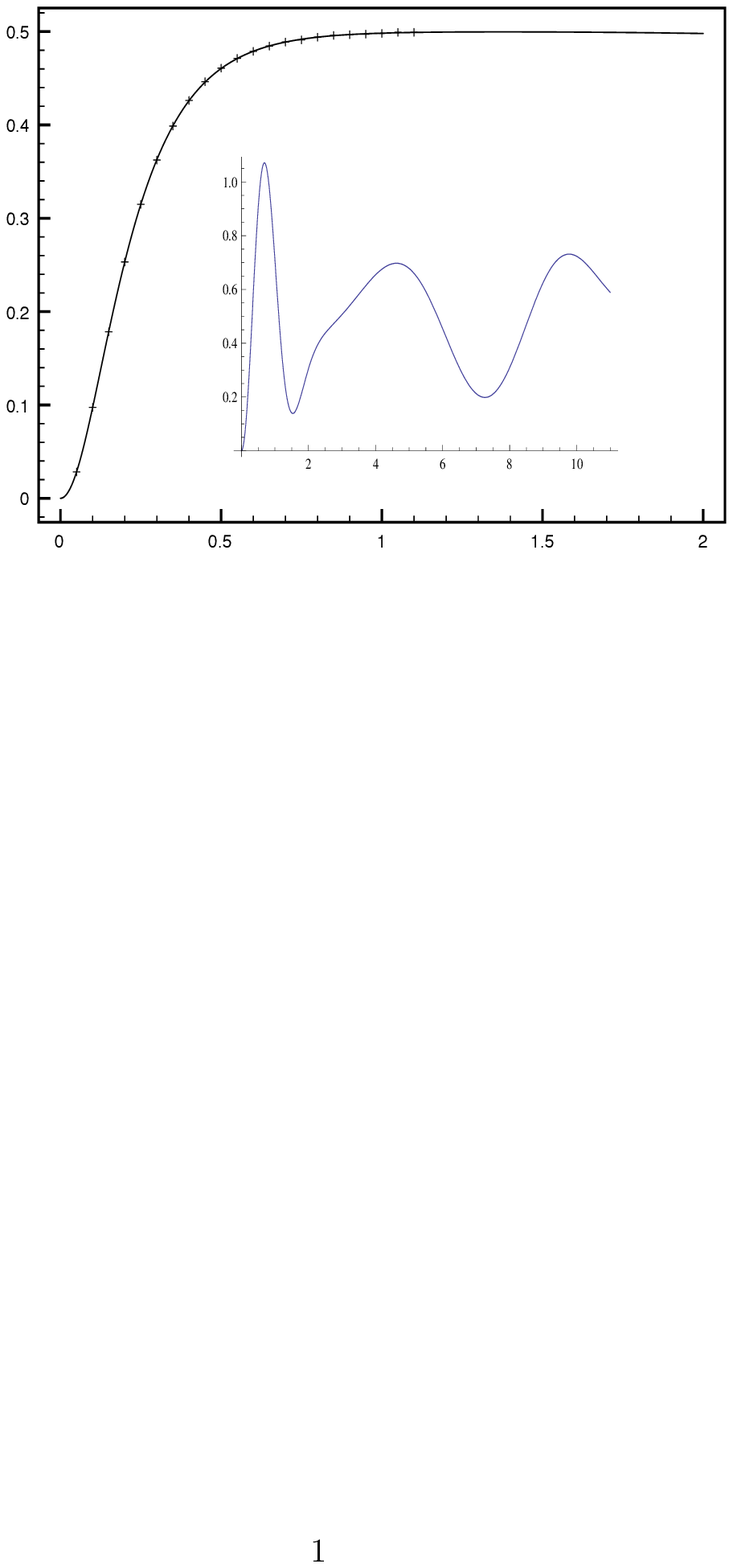}
\end{center}
\vspace{-2mm} 

\caption{\small Results from the 1d Langevin simulation at 
$\th=0,\lb=1$ for $m_4(t)$ versus the Borel transform $M_4(t)$. 
The insert shows the Borel sum $b_4(s)$.} 
\end{figure}

Finally, the result (\ref{borel8}) is also consistent with the recursion 
relation (\ref{borel2}) remaining valid for all $t$ with the 
$m_p(t) = \bra x^p\ket_{\rho_t}$ interpreted as the Borel 
transform $M_p(t)$. Stationary distributions have 
$\dd_t^k m_2 =0,\,k \geq 1,$ and (\ref{borel2}) simplifies to 
$4 \alpha m_p(\infty) = (p-3) m_{p-4}(\infty)$, which is readily solved:
\ba
\label{QL9}
m_p(\infty) \is 
\frac{\Gamma\big( \frac{p+1}{4} \big) }{\Gamma\big( \frac{1}{4} \big) } 
\alpha^{-p/4} \,, \bspace\;\;\, p = 0\! \mod 4\,,
\nonum
m_p(\infty) \is 
\frac{\Gamma\big( \frac{p+1}{4} \big) }{\Gamma\big( \frac{3}{4} \big) } 
m_2(\infty) \,\alpha^{(2-p)/4} \,,\quad \;\;p = 2 \!\mod 4\,.
\ea
This holds irrespective of the validity of the conjecture 
(\ref{pkc}) or its $t \ra \infty$ limiting form (\ref{APKC}). 
Assuming that 
the $\bra \cO \ket_{\rho_t}$ evolution equation in (\ref{oflow2}) 
remains valid 
for all $t$ and $p$ the asymptotic values must obey (\ref{QL9}).
The limit $m_2(\infty)$ remains undetermined in agreement 
with the structure in (\ref{borel3}). A specific choice 
for $m_2(\infty)$ will render (\ref{QL9}) compatible with the 
asymptotic form of (\ref{PKC}): the limit 
$\lim_{\ra \infty} \bra x^p \ket_{\rho_t} = 
\bra x^p \ket_{e^{-S}}$ is  of course trivially evaluated and gives 
\be 
\label{QL10}
\bra x^p \ket_{e^{-S}} = 
\frac{\Gamma\big( \frac{p+1}{4} \big) }{\Gamma\big( \frac{1}{4} \big) } 
\alpha^{-p/4}\,, \quad p = 0\! \mod 2  \,,\quad {\rm Re} \alpha >0\,.
\ee
Numerically, $\bra x^2 \ket_{e^{-S}} = 0.47798\,\lb^{-1/4}$, 
$\bra x^4 \ket_{e^{-S}} = 0.5\,\lb^{-1/4}$,  
$\bra x^6 \ket_{e^{-S}} = 0.7169\,\lb^{-3/4}$. 
One sees that (\ref{QL10}) matches (\ref{QL9}) iff 
$m_2(\infty) = \alpha^{-1/4}\, \Gamma(3/4)/\Gamma(1/4)$. 
Implicitly therefore, the Borel resummation described before 
fixes the parameter $m_2(\infty)$ undetermined by (\ref{QL9}) 
to precisely this value. 

In the $m_p(t)= \bra z^p \ket_{R_t}$ interpretation of the 
moments the flow equation for $\bra \cO\ket_{R_t}$ in (\ref{oflow2}) 
might likewise be valid for all $t$ but with a $m_2(\infty)$ 
value different from the one above. If so, at least 
the $m_p(\infty)$, $p=0\! \mod 4$, and the ratios 
$m_p(\infty)/m_2(\infty)$, $p =2\! \mod 4$, must come out as in 
(\ref{QL9}). In particular $m_4(\infty) =
0.3536\, \lb^{-1/2} e^{-i\th/2}$.

\newsection{Complex Langevin Simulations: Method and Results}

The complex Langevin process underlying (\ref{i6}) is 
realized as a Wiener evolution in the real two dimensional space 
corresponding to the real and imaginary parts of the 
integration variable. Specifically, for a single integration 
variable with complexification $z=x+iy$, one generates an 
ensemble of points in the $(x,y)$ plane by integrating the 
stochastic pair of equations
\ba 
\label{Langpair}
\frac{dx}{dt} &=& -{\rm Re}\bigg(\frac{\partial S(x+iy)}{\partial x}
\bigg) +b_{R}(t)\,,
\nonum
\frac{dy}{dt} &=& -{\rm Im}\bigg(\frac{\partial S(x+iy)}{\partial x}
\bigg) +b_{I}(t)\,,
\ea
where $b_{R},b_{I}$ are independent Wiener (``white noise") processes, with
\ba
\overline{b_{R}(t_1)b_{R}(t_2)} &=& 2A_{R}\delta(t_{1}-t_{2})\,,
\nonum
\overline{b_{I}(t_1)b_{I}(t_2)} &=& 2A_{I}\delta(t_{1}-t_{2})\,,
\ea
and $A_{R}=A_{I}+1$ (complex fluctuation-dissipation relation). 
In practice, we realize this process by discretizing the time:
\ba
\label{Langpairdisc}
x(t+\Delta t) &=& x(t) - 
{\rm Re}\bigg(\frac{\partial S(x+iy)}{\partial x}\bigg)\Delta t  
+ \Delta b_R \,,
\nonum
y(t+\Delta t) &=& y(t) - 
{\rm Im}\bigg(\frac{\partial S(x+iy)}{\partial x}\bigg)\Delta t  
+ \Delta b_I\,,
\ea
where $\Delta b_R$ (resp.~$\Delta b_I$) are Gaussian distributed 
randoms of variance $2A_{R}\Delta t$ (resp. $2A_{I}\Delta t$). Two 
important aspects of our numerical implementation are (see 
also \cite{Aarts2}):
\begin{enumerate}
\item We use continuously (i.e. normally) distributed random 
variables, rather than discrete random step variables, to avoid 
discretization artifacts in the generated ensemble. Thus, our 
ensemble values (starting at the origin $x=y=0$, say) are not
``quantized" but fill out a continuous region in the $x$-$y$ 
configuration space. 
\item More importantly, we have found that the Brownian 
process occasionally wanders into regions where the ``force"
functions $F_{x}={\rm Re}(\frac{\partial S(x+iy)}{\partial x})$, 
$F_{y}= {\rm Im}(\frac{\partial S(x+iy)}{\partial x})$ 
become quite large. To avoid losing accuracy in the discretized 
realization (\ref{Langpairdisc}), we therefore 
readjust the time step $\Delta t$ at every update to ensure 
that the variations $\Delta x,\Delta y$ remain small. In practice, 
this is done by choosing
\be
\label{adapdelt}
\Delta t = \frac{\delta}{1+|F_{x}|+|F_{y}|}
\ee
with the nominal time step $\delta$ chosen at some suitably 
small value (typically, 10$^{-5}$). If one uses a fixed time step, 
we have found that it is quite common to obtain results which 
appear to converge, but are simply incorrect as a consequence 
of rare excursions which distort the result due to a loss of 
accuracy in the discretization of the Wiener process.
\end{enumerate}
 
 We have carried out an exhaustive numerical investigation of 
the Langevin time dependence for both real and complex polynomial 
actions, specifically for the quadratic (``free")  
$S(z) = \om e^{i\theta/2}z^2$, $\om>0\,,0\leq \th <\pi$, detailed in 
Appendix A, and the quartic 
(``interacting") $S(z) = \frac{1}{2} \sqrt{\lambda} 
e^{i\theta/2}\,z^4$, $\lb>0,\,0\leq \th <\pi$, case. 
The Langevin time dependence of the second and fourth moments 
${\mathcal O}(z)=z^2,z^4$ was studied by generating a large
number (typically 1-4 million) of independent trajectories in 
the $x$-$y$ plane (starting at the origin), using the adaptive 
discretized algorithm described above, and terminating each 
trajectory when the desired Langevin time was reached. This 
allowed us to study both the short time and long time asymptotics 
of the complex Langevin process.

For definiteness we present the results for the second moments. 
The behavior of the fourth moments is qualitatively similar
but the breakdown times are often signalled by 
a sudden increase in fluctuations size. 
Figure 4 shows a comparison of the results from the Langevin 
simulation of the second moments with those from the Borel transform
(\ref{borel10}) for several values of $A_I$ and fixed $\th =\pi/2$. 
The statistical errors for the Langevin results are smaller than 
the size of the symbols. The putative equilibrium values for large 
$t$ coincide with those obtained by Aarts \cite{Aarts3}.

\begin{figure}[htb]
\begin{center} 
\includegraphics{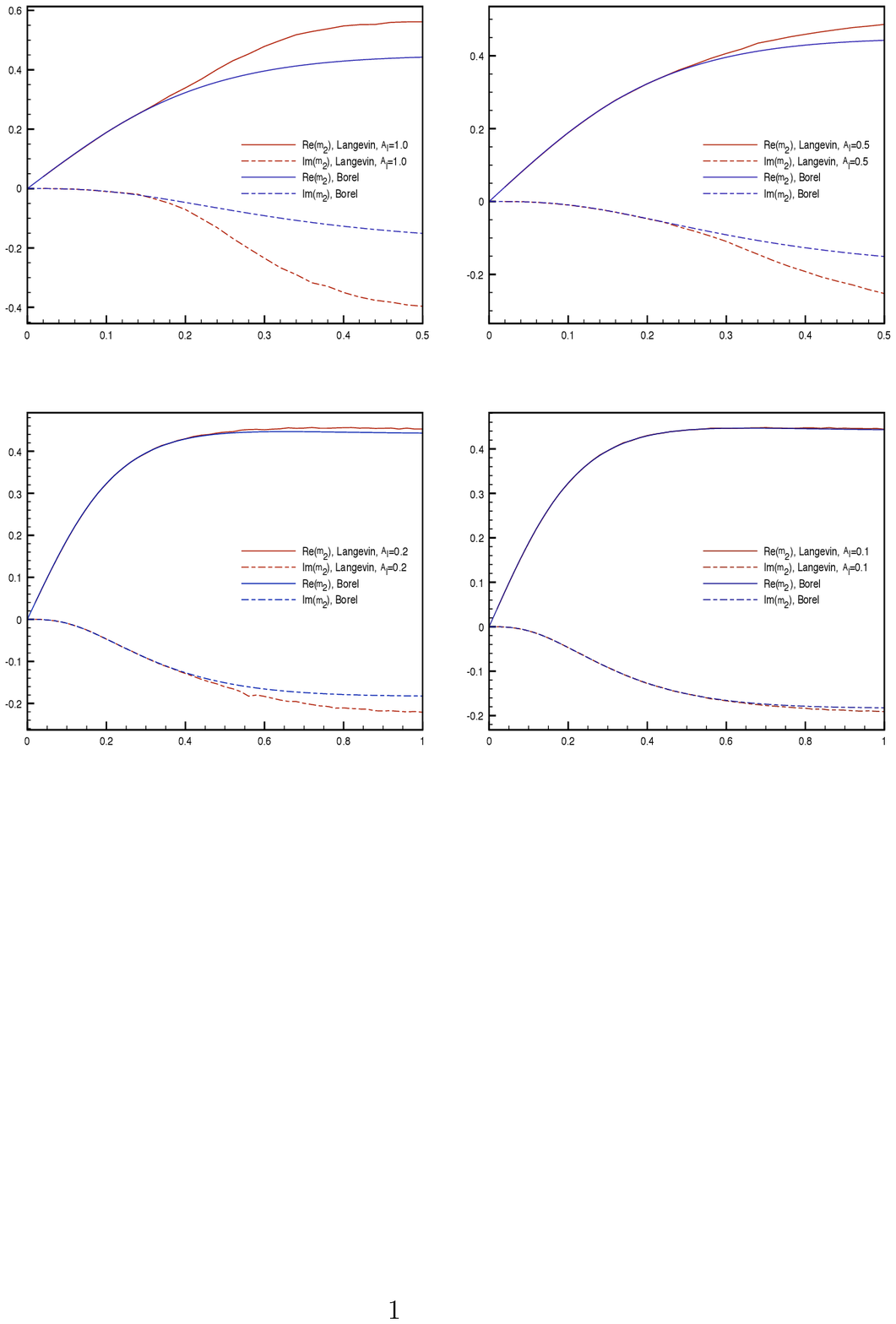} 
\end{center}

\caption{\small Second moments from Langevin simulations (red) for various 
values of $A_I$ and $\th =\pi/2$ compared to the Borel transform (blue).}
\end{figure}
\bigskip

One sees that in each case there is good agreement for 
$0 \leq t \leq t_c(A_I)$ but disagreement for $t > t_c(A_I)$. 
For the time being we define the `breakdown time' $t_c(A_I)$ 
informally as the smallest time at which both results deviate
significantly with respect to the (statistical and systematic) errors.
This rests on the `experimental' fact that the transition 
is relatively sharp; a theoretical understanding of this
phenomenon is currently lacking. Assuming there is a sharp 
transition $t_c(A_I)$ visibly increases with decreasing $A_I$. 
For the values considered one has $t_c \approx 0.16, 0.22, 0.41, 0.67$ 
for $A_I = 1,0.5,0.2,0.1$, respectively. Tentatively this suggests 
a powerlike scaling 
\be 
\label{tcscaling}
t_c(A_I) \sim (A_I + \alpha)^{-\gamma} \,,\quad 
\alpha \approx 0,\;\;\gamma \approx 0.6\,,
\ee
though other functional forms are compatible with the data as well. 
The precise form of the $A_I$ dependence can of course not be pinned 
down numerically and neither can be the limit $\lim_{A_I \ra 0} t_c(A_I)$.

However, the breakdown of the agreement for $A_I>0$ shows that at best 
the refined version (\ref{PKCA}) of the conjecture can hold true. 
A similar conclusion was reached along different lines in \cite{Aarts1}.
In much of the early literature $A_I =0$ was used by default. 
On the other hand the heuristic derivations of (\ref{PKC}) 
do not discriminate between real and complex noise, and neither 
does the framework of \cite{Guralnik1,Guralnik2}. 
  
Although the Langevin results for $A_I>0$ differ for $t>t_c(A_I)$
from the `desired' results, they seem likewise to converge to 
finite asymptotic values $m_p(\infty)|_{\rm Langevin}$. In line with 
the general setting described in Section 3.1 this should equal
the average as computed with a positive equilibrium distribution 
$\vp_0$ in the kernel of $\P$, see (\ref{Pspec5}).   
In the next section we compute $\vp_0$ directly from the defining 
relation $\P \vp_0 =0$ and verify that $\lim_{t \ra \infty} \bra z^p 
\ket_{R_t}  = \bra z^p \ket_{\vp_0}$ indeed holds for the second moments 
considered.

\newsection{Spectrum and ground state of $\P$} 

As stressed before the non-naive action of the semi-group generated by 
the real Fokker-Planck operator $\P$ on holomorphic functions
is the likely culprit for the failure of the method. 
Of course one has very little analytic control over the 
semigroup in question, so one might doubt that 
$\P$ generates a well-defined semigroup at all. 
In the following we present numerical results indicating 
that at least several necessary conditions for this to be 
the case are satisfied. Along the way we also obtain 
approximations to the ground state wave function of $\P$ which   
allows us to independently compute the relevant moments. 

We continue to treat the anharmonic case as a paradigm, 
$S(z) = \frac{1}{2} \sqrt{\lb} e^{i\th/2} z^4$. The real 
Fokker-Planck operator $\P$ in this case reads 
\ba
\label{QL1}
\P \is A_R \dd_x^2 + A_I \dd_y^2 - F_x \dd_x - F_y \dd_y - 
\dd_x F_x - \dd_y F_y\,,
\nonum
F_x \is -2 \sqrt{\lb}\Big[ (x^3 - 3 x y^2) \cos \frac{\th}{2}
+ (-3 x^2 y + y^3) \sin \frac{\th}{2} \Big]\,,
\nonum
F_y \is -2 \sqrt{\lb}\Big[ (3 x^2 y- y^3) \cos \frac{\th}{2} 
+ (x^3 -3 x y^2) \sin \frac{\th}{2} \Big]\,,
\ea   
with $\dd_x F_y + \dd_y F_x =0$. The operator has two manifest symmetries
\be
\label{QL2} 
\P|_{x \ra \lb^{-1/8}x, y \ra \lb^{-1/8}y} = \lb^{1/4} \P|_{\lb =1}\,,
\quad 
\P|_{x \ra -x, y \ra -y} = \P\,,
\ee
which constrain the structure of the spectrum and the eigenfunctions. 
The first one implies that the exact eigenvalues of $\P$ scale like 
$\lb^{1/4}$, just as they do for the complex Fokker-Planck 
operator $\bP$, which in turn follows from (\ref{snorm6}) and the fact 
that $\bH$ and $\bP$ are similar. Since $\P$ is real for all 
parameter values the complex conjugate of an eigenfunction will
again be an eigenfunction with the complex conjugate spectral 
value. This means the spectrum of $\P$ lies symmetric to the real 
axis. It is plausibly purely discrete and the numerical 
results presented below are compatible with the 
structure (\ref{Pspec}), where $\lb^{-1/4} E_n$ is $\lambda$-independent 
and $E_0$ is non-degenerate. Concerning the eigenfunctions $\P \varphi_n = 
E_n \varphi_n$, the scaling relation entails that they can  
be written as a $\lb$-independent function evaluated at 
arguments $\lb^{1/8} x, \lb^{1/8} y$, 
\be 
\label{QL4} 
\vp_n(x,y) = \vp_n|_{\lb =1}(\lb^{1/8} x, \lb^{1/8} y)\,.
\ee
The reflection symmetry $\P|_{x \ra -x, y \ra -y} = \P$ entails 
that the eigenfunctions can be chosen to have definite parity, 
$\vp_n(-x,-y) = \delta_n \vp_n(x,y)$, $\delta_n \in \{\pm 1\}$.
Since we expect the ground state to be positive it must be 
invariant
\be 
\label{QPKC}
\varphi_0(x,y) = \varphi_0(-x,-y)\,,
\ee
which also ensures that odd observables $\cO(x+iy)$ have 
vanishing expectation value (\ref{Aavgs}).


\newsubsection{Spectrum} 

In a first step we now investigate the spectrum of $\P$ 
using a variant of the matrix truncation technique from 
Section 2. Anticipating that zero lies in the spectrum 
the associated eigenvector can be used to approximate 
$\varphi_0$ and to compute the required averages. 
To this end it is useful to choose real basis functions 
like the products $H_k(x) H_l(y)$ of Hermite 
functions in $x$ and $y$. Reality of the approximate $\varphi_0$ 
can then be seen directly and only the test for positivity requires 
further analysis. The price to pay is that the matrix elements
in a real basis are more complicated than in a complex basis 
adapted to (\ref{L5}). We introduce a pair of oscillators 
$[a,a^*]=1=[b,b^*]$ by 
\ba
\label{QL11}
&& x = \frac{1}{\sqrt{2 \om}}(a^* + a) \,,\quad 
p_x = i \sqrt{\frac{\om}{2}} (a^* -a) = - i \dd_x \,, 
\nonum
&& y = \frac{1}{\sqrt{2 \om}}(b^* + b) \,,\quad 
\;p_y = i \sqrt{\frac{\om}{2}} (b^* -b) = - i \dd_y\,, 
\ea 
with $\om := \sqrt{6} \lb^{1/4}$. Further, we decompose $\P$ 
as follows
\ba 
\label{QL12}
&& -\P = A_R \,p_x^2 + A_I \,p_y^2 + 2 \om^2 
\big[(y^2 - x^2) \cos \frac{\th}{2} 
+ 2 x y \sin \frac{\th}{2} \big] 
\\[2mm]
&& + \frac{\om}{12} \cos \frac{\th}{2} [X(x,y) - X(y,x)] 
+ \frac{\om}{12} \sin \frac{\th}{2} [Y(x,y) + Y(y,x)]\,, 
\nonumber
\ea 
where 
\be 
X(x,y) = -i 4 \om (x^3 - 3 x y^2) p_x \,,
\quad 
Y(x,y) = -i 4 \om (x^3 - 3 x y^2) p_y \,.
\ee
The symmetry (\ref{QL2}) now amounts to $\P|_{x \ra x/\sqrt{\om},  
y \ra y/\sqrt{\om}} = \om \,\P|_{\om =1}$, so that $\om$ drops 
out upon insertion of (\ref{QL11}). One finds 
in terms of the oscillators 
\ba 
\label{QL13}
&& -\frac{2}{\om} \P = \big(A_R \!- \!2 \cos \frac{\th}{2} \big) 
(2 a^* a \!+ \!1) + \big(\!\!A_I \!+ \!2 \cos \frac{\th}{2} \big)
(2 b^* b \!+\!1) 
\nonum
&& - \big(A_R\! + \!2 \cos \frac{\th}{2} \big)({a^*}^2 \!+\! a^2) 
+  \big(\!-A_I\!\! + \!2 \cos \frac{\th}{2} \big)({b^*}^2 \!+\! b^2) 
+ 4 \sin \frac{\th}{2} (a^* + a)(b^* + b) 
\nonum
&& + \frac{1}{6} \cos \frac{\th}{2} \,[X(a,b) - X(b,a)] 
+  \frac{1}{6} \sin \frac{\th}{2} \,[Y(a,b) + Y(b,a)] \,,
\nonumber
\ea   
where 
\ba 
\label{QL14} 
X(a,b) \is 
{a^*}^4 \!- \!a^4 \!+ \!2 {a^*}^3 a \!- \!2 a^* a^3 \!+ \!6 {a^*}^2 
\!+ \!6 a^* a \!+ \!3 
-3( {b^*}^2 \!+ \!2 b^* b \!+ \!b^2 \!+\!1)({a^*}^2 \!- \!a^2 \!+\!1)\,,
\nonum
Y(a,b) \is ({a^*}^3 \!+ \!3 {a^*}^2 a \!+ \!3 a^* a^2 \!+\!a^3 \!
+ \!3 a^* \!+ \!3 a)( b^*\! -\!b) 
\nonum
&& - 3(a^* \!+\!a) ({b^*}^3 \!+ \!{b^*}^2 b \!-\! b^* b^2 \!- \!b^3 
\!+ \!3 b^* \!+ \! b)\,.
\ea
The basis of Hermite functions in $\sqrt{\om} x$, $\sqrt{\om} y$ 
corresponds to 
\be
\label{QL15}  
|m,n\ket := \frac{1}{\sqrt{m! n!}} {a^*}^m {b^*}^n |0\ket \,,
\quad a|0\ket = b|0\ket =0\,.
\ee
Finally the matrix elements of (\ref{QL13}) come out as 
\ba 
\label{QL16}
&& \nspace -\frac{2}{\om} \bra kl|\P |mn\ket = 
\Big[\big(A_R - 2 \cos \frac{\th}{2}\big) (2 k+1) +
\big(A_I + 2 \cos \frac{\th}{2}\big) (2 l+1) \Big] \delta_{km}\delta_{ln} 
\nonum
&& - \big( A_R + 2 \cos \frac{\th}{2} \big) 
[ \sqrt{k(k-1)} \,\delta_{k-2,m} + \sqrt{m(m-1)} \,\delta_{m-2,k}]\, 
\delta_{l,n} 
\nonum
&& + \big(\! -A_I + 2 \cos \frac{\th}{2} \big) 
[ \sqrt{l(l-1)} \,\delta_{l-2,n} + \sqrt{n(n-1)} \,\delta_{n-2,l}]\, 
\delta_{k,m} 
\nonum
&& + 4 \sin \frac{\th}{2} 
\big( \sqrt{k} \,\delta_{k-1,m} + \sqrt{m} \,\delta_{m-1,k} \big) 
\big( \sqrt{l} \,\delta_{l-1,n} + \sqrt{n} \,\delta_{n-1,l} \big) 
\nonum
&& + \frac{1}{6} \cos \frac{\th}{2}\, 
[ X_{kl,mn} - X_{lk,nm}]  
+ \frac{1}{6} \sin \frac{\th}{2} \,
[ Y_{kl,mn} + Y_{lk,nm}]\,,
\ea 
where 
\begin{subeqnarray} 
\label{QL17}
 X_{kl,mn} \is 
\delta_{l,n} \sqrt{\frac{k!}{m!}} [ \delta_{k,m+4} + 2(k-2) \delta_{k,m+2} ]
-\delta_{l,n} \sqrt{\frac{m!}{k!}} [ \delta_{m,k+4} + 2(m-2) \delta_{m,k+2} ]
\nonum
&+& \delta_{ln} [ 6\sqrt{k(k-1)} \delta_{k,m+2} + (6k+3) \delta_{k,m}]
\nonum
&-& 3[ \sqrt{(l+2)(l+1)} \delta_{l+2,n} + \sqrt{l(l-1)} \delta_{l-2,n} 
+ (2 l+1) \delta_{l,n}]
\nonum
&\times & 
[ \sqrt{(m+2)(m+1)} \delta_{k,m+2} - \sqrt{m(m-1)} \delta_{k,m-2} 
+ \delta_{k,m}]  
\\[2mm]
Y_{kl,mn} \is [\sqrt{l} \delta_{l-1,n} - \sqrt{l+1} \delta_{l+1,n}]
\,[ \sqrt{(m+3)(m+2)(m+1)} \delta_{m+3,k} 
\nonum
&+ & 3 \sqrt{m+1}(m+1) \delta_{m+1,k} 
+ 3 \sqrt{m} m \delta_{m-1,k} + \sqrt{m(m-1)(m-2)} \delta_{m-3,k}]
\nonum
&-& 3 [\sqrt{k} \delta_{k-1,m} + \sqrt{k+1} \delta_{k+1,m}] 
\,[ \sqrt{(n+3)(n+2)(n+1)} \delta_{l,n+3} 
\nonum
&+ &\sqrt{n+1} (n+3) \delta_{l,n+1} 
- \sqrt{n} (n-2) \delta_{l,n-1} - \sqrt{n(n-1)(n-2)} \delta_{l,n-3}]\,. 
\end{subeqnarray} 
The matrix elements of the transpose operator 
$\P^T = A_R \dd_x^2 + A_I \dd_y^2 + F_x \dd_x + F_y \dd_y$,
are obtained similarily and read 
\ba 
\label{QL18}
&& \nspace -\frac{2}{\om} \bra kl|\P^T |mn\ket = 
[A_R (2 k+1) + A_I(2 l+1)] \delta_{km}\delta_{ln} 
\nonum
&& - A_R 
[ \sqrt{k(k-1)} \,\delta_{k-2,m} + \sqrt{m(m-1)} \,\delta_{m-2,k}]\, 
\delta_{l,n} 
\nonum
&& -A_I 
[ \sqrt{l(l-1)} \,\delta_{l-2,n} + \sqrt{n(n-1)} \,\delta_{n-2,l}]\, 
\delta_{k,m} 
\nonum
&& - \frac{1}{6} \cos \frac{\th}{2}\, 
[ X_{kl,mn} - X_{lk,nm}]  
- \frac{1}{6} \sin \frac{\th}{2} \,
[ Y_{kl,mn} + Y_{lk,nm}]\,.
\ea 
One can check 
\be 
\label{QL19} 
\bra kl|\P^T |mn\ket = \bra mn|\P |kl\ket\,,
\ee
so for the matrix trunctions the last spectral equality 
in (\ref{Pspec2}) is manifestly satisfied. 

The numerical computation of the spectrum is now straightforward.  
After conversion of the double index $k,l =0, \ldots ,N-1$ into a 
single via $i = k N + l+1$, the resulting matrix $P_{ij} := 
-(2/\om) \bra kl|\P|mn\ket$, $i,j =1, \ldots ,N^2$, 
can be diagonalized numerically. The reliability and accuracy of 
the approximate eigenvalues can be assesed by increasing $N$. 
Since some of the eigenvalues are complex the ordering is 
done by modulus, i.e.~$n>m$ if $|E_n| > |E_m|$. With a 
$C$ code implementing the Hessenberg transformation running 
times are $1-6$ hours for $N=80 -100$ on a conventional laptop 
with effectively $2G$ RAM usage. In order to reach $N=160$ without 
cumbersome recoding we used a $16G$ RAM workstation with 
running times $8-140$ hours. The approximate spectra stabilize 
convincingly at the $O(10^{-3})$  level, from the trend
upon varying $N$ between $100-160$ we guesstimated a systematic 
error. For illustration we report the results for 
a specific case: $A_I =1, \th = 0,\pi/2$ in Table 4. By the 
argument after Eq.~(\ref{Pspec3}) the lowest eigenvalue must 
be zero.  
\bigskip

\begin{table}[htb]
\centering

\begin{tabular}{|c|c|| c|}
\hline
$n$ &  $A_R =2, A_I =1, \th =0$ &$A_R =2, A_I =1, \th =\pi/2$
\\ 
\hline\\[-4.5mm]
$0$ & $0.002(2)$                  & $0.006(6)$
\\\hline                          
$1$ & $2.0725(2)$                  & $1.921(2)$
\\\hline
$2$ & $5.4920(2)$                  & $5.833(2)$
\\\hline
$3$ & $8.5377(2)$                  & $7.885(2)$
\\\hline                       
$4$ & $8.314(3) \pm i\, 5.798(3)$   &  $8.590(3) \pm i\, 6.092(3)$
\\\hline
$5$ & $13.962(3) \pm i\, 5.998(3)$  &  $13.125(3) \pm i\, 5.254(3)$
\\\hline                         
$6$ & $14.310(3) \pm i\, 8.346(3)$  & $15.074(3) \pm i\, 9.160(3)$
\\\hline
\end{tabular}
\caption{\small Low lying eigenvalues for $\P$ with $\lb =1$ based 
on $N=100-160$ truncations.} 
\end{table}

One sees that the first few eigenvalues are real while higher 
excited states typically come in complex conjugate pairs. 
Importantly the real parts are nonnegative so that the 
spectrum is compatible with $\P$ being the generator 
of a semigroup with real kernel. The spectrum for other 
parameter values was found to be qualitatively similar and is 
compatible with (\ref{Pspec}). Positivity of the kernel
is essential for the interpretation of $R_t$ as a probability 
measure but is not manifest from the spectrum.

\newsubsection{Ground state} 

In contrast to the quadratic case the determination of the 
ground state wave function is nontrivial. First note 
that  now $\exp\{-S(x+iy)\}$ is not even an eigenfunction of $\P$. 
This is because although $\dd_y$ acts like $i \dd_x$ on holomorphic 
functions, the replacement is illegitimate in the $\dd_y F_y$ 
term. This results in $e^{+S} \P e^{-S} = - i {\rm Im} \dd_x^2S 
- \dd_y F_y = 6 \sqrt{\lb} e^{-i\th/2} (x - i y)^2$.    
The special real solution (\ref{OL5}) generalizes to the 
interacting case but is again not integrable. For generic $A_R$ it 
is tempting to search for elements in the kernel of the 
form 
\be 
\label{QL20}
\varphi_0(x,y) = \exp\{ - A_0 x^4 - 3B_0 x^3 y - 6 C_0 x^2 y^2 
- 3 D_0 x y^3 - E_0 y^4\} \,.
\ee
By direct computation one sees that no such solution exists 
for $A_R \neq 1/2$. Since 
$\P \varphi_0 = E_0 \varphi_0$ is an elliptic second order differential 
equation one expects the existence of a solution to be 
determined by general principles. The usual existence 
theorems however refer to bounded domains with Dirichlet 
boundary conditions. In the situation at hand compactification of 
$\R^2$ to the unit square, say, introduces singularities 
in the coefficient functions of $\P$ towards the boundary 
where the solution is supposed to vanish. In the interior the 
desired solution must be non-negative. Establishing 
(non-)existence of a solution along these lines 
therefore may be non-trivial.

Using the matrix truncation technique the numerical determination 
of the ground state is however straightforward. Once the ground 
state energy $E_0$ is reliably identified at the chosen 
truncation level $N$ one can compute the associated eigenvector 
of $P_{ij}$ to high accuracy by iterated action of the resolvent. 
The resulting vector 
$v_i,\, i =1, \ldots, N^2$ is converted into the 
approximate ground state wave function by the transformation 
\be 
\label{QL21}
\vp_0(x,y) = \sum_{i=1}^{N^2} v_i \, H_{k(i)}(\sqrt{\om} x) 
H_{l(i)}(\sqrt{\om} y) \,,
\ee
where the $\om$-dependence is restored 
and $k(i) = (i-1 - {\rm mod}(i-1,N))/N$, $l(i) = {\rm mod}(i-1,N)$ 
converts the indices, with ${\rm mod}(m,n)$ defined as $m$ modulo $n$. 
The required basis of Hermite functions is conveniently programmed
using the recursion relation. As a test on the accuracy of the
conversion one can check $\int\!dx dy\, \vp_0(x,y)^2 = 
\om^{-1} \sum_i v_i^2$.

We computed the eigenvector $v_i, \, i=1,\ldots N^2$, 
for the same parameters sets as in Table 4 and converted 
it into the corresponding ground state wave function $\varphi_0$ 
using (\ref{QL21}). The results are shown in Figure 6. 
It is numerically nontrivial that they come out strictly positive
and therefore can serve as a probability measure.

\begin{figure}[htb]
\vspace{5mm} 

\begin{center} 
\includegraphics{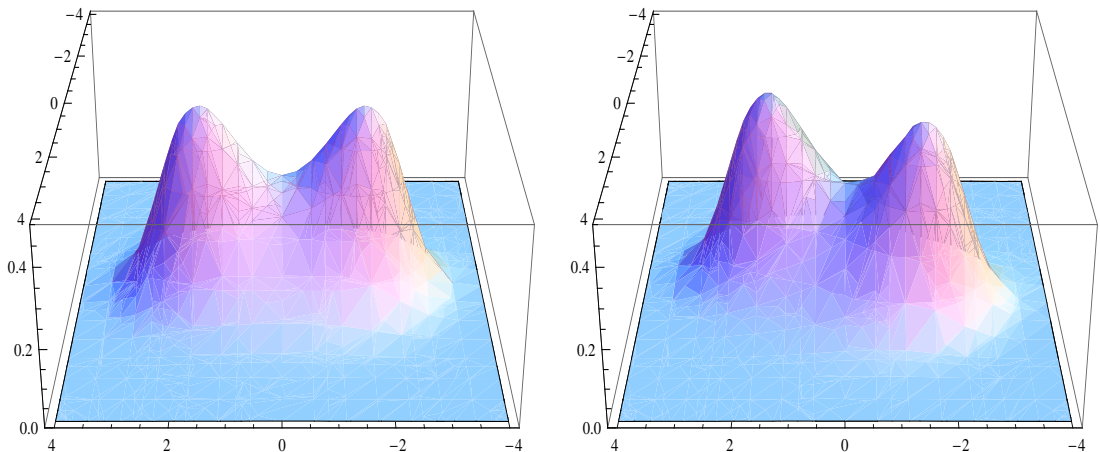} 
\end{center} 
\vspace{-3mm}

\caption{\small Ground state wave function for $\P$ for truncation 
$N=140$ and parameters $A_R =2, A_I =1$. Left $\th =0$, right $\th =\pi/2$.}
\end{figure}

For nonzero $\th$ one sees that the principal axes in Figure 
5 is rotated by an angle $\th/4$. By analogy with the 
non-interacting case described in Appendix B one should not 
expect, however,  that $\vp_0$ differs from $\vp_0|_{\th =0}$ only 
by a rigid rotation. For other values of $A_I$ the results for 
the ground state wave functions are qualitatively similar: 
a double peak is visible whose principal axes is rotated by 
an angle $\th/4$. The main effect of decreasing $A_I$ is to 
shrink the extension of the level surfaces in the $y$-direction
and to raise the summits. This is illustated in Figure 6.

\begin{figure}[htb]
\begin{center} 
\includegraphics{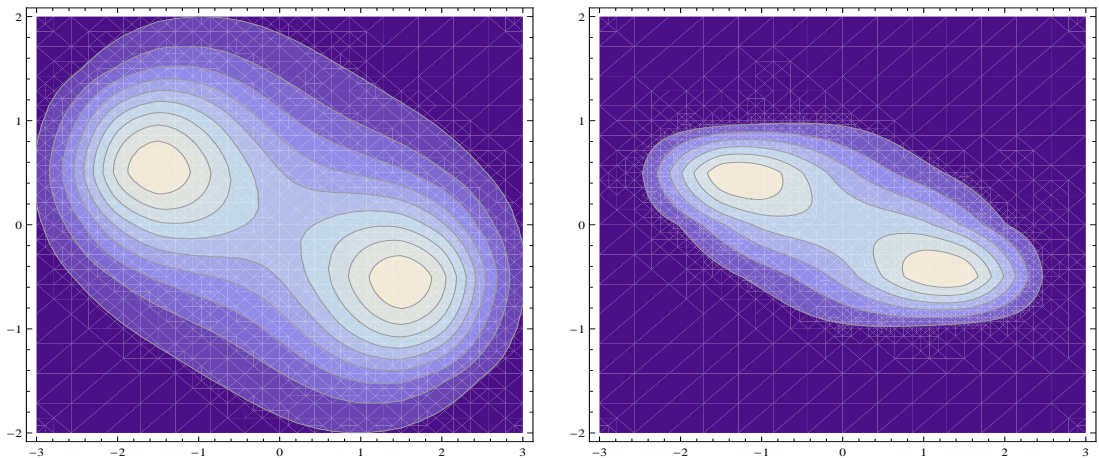}
\end{center} 
\vspace{-3mm} 

\caption{\small Level surfaces of $N=140$ ground state wave functions 
for $\th =\pi/2$. Left $A_I =1$, right $A_I =0.1$.}
\end{figure}

Finally we computed the second  moments for the same 
parameter values as in the Langevin study. The Langevin 
moments for large times have previously been obtained by 
Aarts \cite{Aarts3}; ours are in full agreement. Here 
we see that the latter also agree within the errors 
with those obtained from the directly computed putative 
equilibrium measure $\vp_0$.

\begin{table}[htb]
\centering

\begin{tabular}{|c|c|c|c|c|}
\hline
$A_I$ & $1$  & $0.5$ & $0.2$ & $0.1$
\\\hline
$m_2(\vp_0)$ & $0.54 - i 0.41$ & 
$0.48 - i 0.30$ & $0.45 - i 0.23$ & $0.44 - i 0.20$ 
\\
$m_2(\infty)_L$ & $0.56 -i 0.41$ & 
$0.49 -i 0.29$ & $0.45 - i0.23$ & $0.44 - i 0.20$ 
\\\hline
\end{tabular}
\caption{\small Asymptotic second moments for $\th = \pi/2$ as computed 
from the $N=150$ ground state $\vp_0$ and from the Langevin dynamics. 
Both agree within the errors but differ from the `desired' one set by 
the complex measure: $m_2(\infty) = 0.441596 -i 0.182915$.} 
\end{table}

Hence 
\be 
\label{Llimit}
\lim_{t \ra \infty} \bra z^p \ket_{R_t}  = \bra z^p \ket_{\vp_0} \,,
\quad \P \vp_0 =0\,,\;\;\vp_0 >0\,,
\ee 
holds for the second moments considered. This provides direct 
computational evidence that the interplay between the stochastic 
differential equations (\ref{Langpair}) and the semigroup generated 
by the real Fokker-Planck operator $\P$ is indeed as expected.
Further the $t \ra \infty$  limit is compatible with the projection
property onto the ground 
state $\vp_0$,  as stipulated in (\ref{Pspec5}). Nevertheless 
(\ref{APKC}) fails for $A_I>0$ which via (\ref{oflow7}) signals 
that property (iv) in Section 3.2 fails.

\newpage
\newsection{Conclusions} 

We critically reexamined the complex Langevin method in the 
paradigmatic case of a $e^{i\th/2} \phi^4$ complex measure. 
In contrast to earlier studies we focussed on the temporal 
rather than the equilibrium aspects of the stochastic dynamics. 
Our main result is that the short time asymptotics encapsulates
information also about the equilibrium aspects. Both sides 
of the conjectured identity (\ref{PKC}) have identical 
asymptotic expansions around $t=0$. The coefficients are such 
that the series is Borel summable and the Borel transform
$M_p(t)$ of the moments' time series correctly captures 
the dynamics of the complex measure: $M_p(t) = \bra x^p \ket_{\rho_t}$,
for all $t \geq 0$. The resummation of the $t=0$ asymptotic expansion 
therefore provides a practically usable alternative to the 
complex Langevin method, at least for low dimensional systems. 

As to the validity of the conjecture itself, the 
results of sections 4 and 5 provide a counterexample
for its validity for $A_I >0$, in line with 
\cite{Aarts3} and similar results in other systems \cite{Aarts2}. 
At best the modified form (\ref{PKCA}) of the conjectured 
identity can therefore hold true. Even for this variant, however, 
counterexamples can be found. For example, for $S(x) = 
\frac{1}{2} \sqrt{\lb} e^{i\th/2} (x^2 \pm \om^2)^2$ we find 
for $A_I=0$ agreement of the putative equilibrum Langevin
results with the analytical answer when the potential 
has a single well (plus sign) but not for a double well 
(minus sign) and $\th \neq 0$. See \cite{Klauder2} for related 
numerical results. In the $\om^2=0$ case considered here the detailed 
analysis of the non-selfadjoint generators $\bH$ and $\P$ 
(spectrum, spectral expansion, ground state) 
revealed no pathological features of the semigroups 
$e^{- t \bH},\,t>0$, and $e^{t \P},\,t>0$. In particular
both semigroups project for $t \ra \infty$ correctly 
only the respective ground states, $e^{-S/2}$ and $\vp_0$. 
For $e^{- t \bH},\,t>0$, the results of Section 2 come 
close to a mathematical proof thereof. The deviation 
from selfadjointness produces nontrivial spectral norms
which impact the quasi-spectra and limit the extension 
to a holomorphic semigroup. For $t$ real and large 
relevant in the present context, however, 
the $(e^{t \bP})(x,x')$ convolutions converge to $e^{-S}$ averages 
even for couplings where the potential is unbounded from below. 
For the $e^{t \P},\,t>0$, semigroup analytical control
over the projection property is more difficult, but we regard 
the computational evidence in Section 5 as quite convincing.

Although both sides of  (\ref{PKC}) have well defined limits,
for $A_I >0$ the limits are different: 
\be 
\bra x^P\ket_{e^{-S}} = \lim_{t \ra \infty} \bra x^p\ket \neq 
\lim_{t \ra \infty} \bra z^p\ket_{R_t} = 
\bra z^p \ket_{\vp_0} \,.
\ee 
We isolated as the culprit the limited validity 
of the identity $\int \!dx' dy' \,\exp(t \P^T)(x,0;x',y') 
\newline \cO(x'+iy') 
= \int\! dx' \,\exp(t \bP^T)(x,x') \cO(x')$, for all $t >0$ and 
all $x$. The identity holds by construction pointwise in $x$ 
as an asymptotic series in $t$ and both sides are termwise 
independent of $A_I$.  For $t \ra \infty$ the identity turns into
(\ref{oflow7}) which fails for $A_I>0$. It does so not because 
an integrations by part step fails but because the ground state 
$\vp_0$ has the wrong structure. Since $\vp_0$ is manifestly 
$A_I$ dependent the $A_I$-independence of averages of holomorphic 
observables would be a nontrivial bonus property -- which 
$\vp_0$ simply fails to have. 

Another way to look at the problem is in terms of the 
under-determination of the observable flow (\ref{oflow2}). 
The approach of \cite{Guralnik1,Guralnik2} aims at selecting 
the solution of the stationary equations corresponding to 
the complex Langevin process driven by $\P$ via minimization 
of a suitable effective potential. However the one-parametric 
family of $\P$'s used will in general not produce results 
independent of the parameter. The parameter drops out in the 
short time asymptotics which provides one rationale for the 
temporal Borel resummation. When applicable, the Borel resummation 
augments the missing piece of information in a way compatible 
with certain analyticity properties in $t$ and the coupling. 

Compared to the field theoretical setting aimed at we 
ignored mass and kinetic terms. In line with earlier 
investigations \cite{Klauder2,Gausterer,Guralnik1,Guralnik2,Bernard}
we regarded the analysis of a pure potential interaction 
as crucial. The inclusion of subleading mass and kinetic 
terms should not affect the qualitative aspects of the 
picture obtained. In particular it should be interesting 
to see whether the temporal Borel resummation technique extends to 
other systems. 
\vspace{1cm}

{\tt Acknowledgements:} We are indebted to G.~Aarts for
communicating his Langevin simulation results on the $A_I$ dependence 
of the equilibrium moments for the quartic action before 
publication \cite{Aarts3}. A.D. also acknowledges useful conversations 
with E.~Seiler and the hospitality of the Max-Planck-Institut,
Munich. A.D. is partially supported by the National Science 
Foundation through grant PHY-0854782. M.N. is on leave 
of absence from the CNRS, France, and acknowledges partial
support by the visitor program of the Department of Physics
and Astronomy of the University of Pittsburgh. Computational 
resources were provided by the ``Frank'' high performance computing cluster 
of the Center for Simulation and Modeling at the 
University of Pittsburgh.


\newpage

\appendix

\newpage
\newappendix{The Langevin method for quadratic complex actions} 
\setcounter{equation}{0}

In the noninteracting case most of the constituents 
of the complex Langevin method can be computed explicitly,
although previously only real noise seems to have been 
considered \cite{CLgaussian,Ambjorn}. Here we allow for 
$A_I >0$ and show that the flow equations (\ref{oflow2}) 
uniquely determine the time dependent moments without the 
need of knowing the real and complex propagation kernels. 
The validity of the Parisi-Klauder conjecture for all 
$A_I \geq 0$ follows. The equilibrium distribution 
of the real measure can be obtained in closed form and the 
numerical evaluation of the equilibrium distribution via the 
complex Langevin simulation shows perfect agreement with 
the analytical result. For the complex propagation kernel 
we briefly review Davies' construction and present a 
generating function for the spectral norms.   

We begin by describing the relation between the familiar 
Mehler kernel and the complex transfer operator with its 
nontrivial spectral norms. The hamiltonian is that of the 
complex harmonic oscillator, 
\be
\label{cosc1} 
\bH_{\th} = p^2 + \om^2 e^{i\th} q^2\,,\quad \om>0\,,\;0 \leq \th <\pi\,,
\ee
where the phase is again normalized so that it corresponds 
to the Wick rotation angle of the original problem. 
Note that for $\pi/2 < \th < \pi$ the potential is 
unbounded from below. 
On account of a scaling argument (\ref{cosc1}) has 
spectrum $E_n = \om e^{i \th/2} (1 + 2n),\,n \in \N_0$,
with eigenfunctions $\Omega_n(q) \sim H_n(e^{i\th/4}q)$,
where the Hermite functions $H_n$ are ortho-normalized 
with respect to the $L^2$ inner product. 
The complex eigenfunctions $\Omega_n$ also admit a Fock space 
description in terms of creation and annihilation operators 
\ba
\label{cosc2}
a_{\th} \is  e^{i\th/4} \sqrt{\frac{\om}{2}} q + i 
e^{-i \th/4} \frac{1}{\sqrt{2 \om}} p = \cos \frac{\th}{2} a + 
i \sin \frac{\th}{2} a^*\,,
\nonum
\bar{a}_{\th} \is  e^{i\th/4} \sqrt{\frac{\om}{2}} q - i 
e^{-i \th/4} \frac{1}{\sqrt{2 \om}} p = \cos \frac{\th}{2} a^* + 
i \sin \frac{\th}{2} a\,,
\ea 
where $[a_{\th}, \bar{a}_{\th}] = [a, a^*]=1$. Since 
$a_{\th}^* = \bar{a}_{-\th}$ the $L^2$ adjoint does not 
preserve the commutation relations. Nevertheless 
$\bH_{\th} = \om e^{i\th} (2 \bar{a}_{\th} a_{\th} +1)$ 
holds and $|n) := n!^{-1/2} \bar{a}_{\th}^n |0)$, 
$a_{\th} |0)=0$, are the realization of the eigenfunctions 
$\Omega_n$. Note that the ($\th$-dependent) $|0)$ can be viewed 
as a coherent state over the $\th=0$ Fock vacuum $|0\ket$. 
To define the relevant inner product we introduce 
a linear anti-involution $\iota$ on the algebra $\cA$ generated 
by the $a_{\th}, \bar{a}_{\th}$ by 
\be 
\label{cosc4} 
\iota(a_{\th}) := \bar{a}_{\th} \,, \quad 
\iota(z x) = z \iota(x)\,, \quad \iota(xy) = \iota(y) \iota(x) \,,
\quad z \in \C,\; x,y \in \cA\,.
\ee 
Then $(x|0),y|0))_{\th}  := (|0), \iota(x) y |0))_{\th}$ with 
$(|0), |0))_{\th} =1$ defines a $\R$-bilinear positive quadratic form 
over $\cA$. In particular the Fock space realization of the 
$\Omega_n$ is an orthonormal basis with respect to $(\;,\;)_{\th}$. 
Returning to position space and the usual $L^2$ inner product 
$\bra \psi,\varphi \ket = \int \!dq \, \psi(q)^* \varphi(q)$,
this means the $\Omega_n^*, \Omega_n$ form a bi-orthogonal basis 
in $L^2$ while the quantities $\bra \Omega_n, \Omega_n\ket = 
N_n(\th)$ are the spectral norms, see Eqs.~(\ref{snorm1}) -- 
(\ref{snorm3}). On account of (\ref{cosc2}) they can in principle
be computed algebraically. More conveniently a generating functional 
for the $N_n$ can be obtained from the Mehler kernel. In the present 
conventions it reads 
\ba 
\label{cosc5}
\big( e^{-\frac{t}{2} \bH_0}\big)(q,q') \is 
\frac{1}{\sqrt{2 \pi \sinh t}} 
\exp\Big\{ - \frac{q^2 + {q'}^2} {2 \tanh t} + 
\frac{q q'}{ \sinh t} \Big\}
\nonum
\is \sum_{n \geq 0} e^{-t (1/2 +n) } H_n(q) H_n(q')\,,
\ea   
where we take $\om =1$ from now on. 
Pointwise the substitution $q \mapsto e^{i\th/4} q, 
q' \mapsto e^{i\th/4} q'$ is legitimate producing the 
transfer operator of the complex harmonic oscillator 
\be
\label{cosc6}
\big( e^{-\frac{t}{2} \bH_0}\big)(e^{i\th/4}q,e^{i\th/4}q') =
\sum_{n \geq 0} e^{-t (1/2 +n) } P_n(q,q') = 
\exp\Big\{\!-\! \frac{t}{2} e^{-i\th/2} \bH_{\th}\Big\}(q,q')\,,
\ee
with $P_n(q,q') = \Omega_n(q) \Omega_n(q')$. Assuming that for 
$q=q'$ integration and summation can be exchanged one obtains
\be 
\label{cosc7} 
\bigg( \frac{r}{(r+r^{-1}) \cos \frac{\th}{2} -2} \bigg)^{1/2} 
= \sum_{n \geq 0} r^{-n} N_n\,,\quad r = e^t\,.
\ee
In particular 
\be 
\label{cosc8}
N_0 = \frac{1}{\cos^{1/2} \frac{\th}{2}} \,,\quad 
N_1 = \frac{1}{\cos^{3/2} \frac{\th}{2}} \,,\quad 
N_2 = \frac{5 - \cos \th}{4\cos^{5/2} \frac{\th}{2}} \,,
\quad \mbox{etc}\,, 
\ee
from which one also infers a powerlike divergence as $\th \ra \pi_-$.
In addition one has the asymptotics
\be 
\label{cosc9}
\lim_{n \ra \infty} \frac{1}{n} \ln N_n = \gamma(\th) < \infty\,,
\ee
where an explicit expression for $\gamma(\th)$ is known
\cite{Davies1}. Norm convergence of the sum in 
(\ref{cosc6}) requires that $\frac{t}{2} {\rm Re} E_n - 
\ln N_n$ is negative for large $n$, leading to the conclusion
that 
\be 
\label{cosc14}
t > t_* = \frac{\gamma(\th)}{\cos \frac{\th}{2}} \,,  
\ee
suffices for norm convergence. In particular the 
$t \ra \infty$ limit still projects onto the ground state,
see (\ref{snorm22}). The semigroup $t \ra e^{-t \bH}$ 
in fact defines a bounded holomorphic semigroup for 
certain $\th$-dependent sectors of the complex $t$ plane
\cite{Davies2,Boulton}. In addition (\ref{cosc4}) affects 
resolvent estimates and the quasi-spectra \cite{Davies1,Boulton}.

The hamiltonian (\ref{cosc1}) is up to an additive constant the 
Fokker-Planck hamiltonian for the complex action 
$S(q) = \om e^{i\th/2} q^2$. By construction it is up to a sign 
isospectral to the complex Langevin operator $\bP$, viz 
\ba 
\label{OL0}
\bP \is \dd_x^2 + 2 \om e^{i\th/2} x \dd_x + 2 \om e^{i\th/2}\,,
\nonum
\bH \is -e^{S(x)/2} \bP e^{-S(x)/2} = -\dd_x^2 + \om^2 e^{i\th} x^2 - \om\,,
\ea
where we omit the subscript $\th$ from now on. The real 
Langevin operator reads  
\be 
\label{OL1} 
\P = A_R \dd_x^2 + A_I \dd_y^2 
+ 2\om \Big( x \cos \frac{\th}{2} - y \sin \frac{\th}{2} \Big) \dd_x 
+ 2\om \Big(y \cos \frac{\th}{2} + x \sin \frac{\th}{2} \Big) \dd_y 
+ 4 \om \cos \frac{\th}{2} \,.
\ee
The propagation kernel $t \mapsto e^{t \bP}$ has been studied before
in terms of $\bH$. The propagation kernel $t \mapsto e^{t \P}$ is 
nontrivial even in the free case. Remarkably neither of these 
objects is needed to prove the Parisi-Klauder conjecture 
for the quadratic action.  

The flow equations (\ref{oflow2}) for the powers (\ref{oflow4}) 
now assume the simple form $\dd_t m_p = p(p-1) m_{p-2} 
- 2 \om p m_p$, for $\th=0$, and the dependence on the phase 
can trivially be restored by rescaling $\om \mapsto e^{i\th/2} \om$. 
These are decoupled differential equations with solution 
\be 
\label{OL2}
m_p(t) = e^{- 2 p \om t} m_p(0) + p (p-1) 
\int_0^t \! dt' m_{p-2}(t') e^{2 \om p(t'-t)} \,.
\ee
Since $m_0(t) =1$, all moments are determined recursively 
up to their values at $t=0$. Since by assumption the initial 
values in both interpretations (\ref{oflow4}) coincide, 
all moments coincide for all $t$, thereby verifying the 
Parisi-Klauder conjecture in this case. Alternatively, one 
can use the evolution $\dd_t \cO = (\dd_z^2 - \dd_z \cO \dd_z) \cO$ 
to define `eigen-observables' 
and reach the same conclusion. The eigenvalue equation translates 
into
\be 
(\dd_x^2 - 2 \om x \dd_x)\cO_n = - 2 n \om \cO_n\,,
\ee 
which is the defining relation for the $n$-th Hermite 
polynomial related by $\cO_n = \Omega_n e^{-S/2}$ to the 
Hermite functions $\Omega_n$. The flow equations (\ref{oflow2}) 
are trivially solved
\be 
\label{OL3}
\bra \cO_n \ket_{R_t} = e^{-2 n \om t} \bra \cO_n \ket_{R_0} = 
e^{-2 n \om t} \bra \cO_n \ket_{\rho_0} = \bra \cO_n \ket_{\rho_t}\,,
\ee
and are equivalent to (\ref{OL2}).

Since (\ref{OL2}) verifies the Parisi-Klauder conjecture 
one expects that its $t \ra \infty$ limit verifies the 
asymptotic form of the conjecture. Indeed, for 
$t \ra \infty$ the dependence on the initial values drops 
out in (\ref{OL2}) and the limits are directly related by the 
recursion $2 \om m_p(\infty) = (p-1) m_{p-2}(\infty)$. Hence 
\be 
\label{OL4} 
m_p(\infty) = \frac{p!}{(p/2)!} \Big( \frac{1}{4 \om} \Big)^{p/2}
=\bra x^p \ket_{e^{-S}} = \bra z^p \ket_{\vp_0}\,, 
\ee  
where the last equality follows from the uniqueness of the solution 
(\ref{OL2}), assuming only that $\vp_0$ exists.  

It is instructive to see that the equilibrium distribution $\varphi_0$ 
can in the harmonic case be found explicitly. We begin 
by searching for solutions of $\P \varphi_0 = E_0 \varphi_0$ 
in the form 
\be
\label{OL4a} 
\varphi_0(x,y) = \exp\{- A x^2 - 2 B xy - C y^2\}\,.  
\ee
For generic $A_R = A_I +1$ one finds three solutions. In addition 
to the expected 
\be 
\label{OL5}
\varphi_0(x,y) = \exp -S(x+iy) \,,\;\;E_0 = 2 \om e^{-i \th/2} \,,
\ee
and its complex conjugate, there is a real solution with $E_0 =0$: 
\ba
\label{OL6}
A_0 \is \frac{2 \om \cos \frac{\th}{2} ( 2 \cosh \frac{\alpha}{2} - 1
- \cos \th)}{\cosh \alpha - \cos \th}
\nonum
B_0 \is \frac{\om (\sin \frac{\th}{2} + \sin \frac{3 \th}{2})}%
{\cosh \alpha - \cos \th}
\nonum
C_0 \is \frac{2 \om \cos \frac{\th}{2} ( 2 \cosh \frac{\alpha}{2} + 1
+ \cos \th)}{\cosh \alpha - \cos \th}\,,
\ea
where we set $A_R = \cosh^2 \alpha/4$, $A_I = \sinh^2 \alpha/4$.
For $\alpha =0$ this reduces to the result in \cite{CLgaussian,Ambjorn}. 
The solution becomes non-normalizable for $\th \ra \pi_-$ as  
$\varphi_0(x,y) \ra 1$. After computing the averages, however, 
the limit $\th \ra \pi_-$ is well-defined, see (\ref{OL10}) below.
The coefficients in (\ref{OL6}) are such that 
\ba
\label{OL7} 
&& \left( \begin{array}{cc} A_0 & B_0 \\[2mm]
B_0 & C_0 \end{array} \right) = 
\left( \begin{array}{cc} \cos \frac{\th}{4} & \sin \frac{\th}{4} \\[2mm]
-\sin \frac{\th}{4} & \cos \frac{\th}{4} \end{array} \right) 
\left( \begin{array}{cc} \lb_+ & 0 \\[2mm]
0 & \lb_- \end{array} \right)
\left( \begin{array}{cc} \cos \frac{\th}{4} & -\sin \frac{\th}{4} \\[2mm]
\sin \frac{\th}{4} & \cos \frac{\th}{4} \end{array} \right) 
\nonum
&& \lb_{\pm} = \frac{2 \om \cos \frac{\th}{2} }{\cosh \frac{\alpha}{2} 
\pm  \cos \frac{\th}{2}} \,,\quad \frac{1}{\lb_+} - \frac{1}{\lb_-} = 
\frac{1}{\om}\,.
\ea
Note that the eigenvalues $\lb_{\pm}$ are $\th$-dependent, 
so the $0 < \th < \pi$ distribution differs from the one at $\th=0$ 
not just by a rigid rotation. 

Specifically for $A_R =1/2$, i.e.~$\alpha = i\pi \, \mod \, i2\pi $, 
a second real solution with $E_0 = 0$ exists:
\be 
\label{OL8} 
\varphi_0(x,y) = \exp\{ - S(x+iy) - S^*(x-iy) \}\,,\quad 
A_R = 1/2\,.
\ee
The solutions (\ref{OL5}), (\ref{OL8}) are not in $L^1(\R^2)$ while 
(\ref{OL6}) is integrable for all $0 \leq \th < \pi$, provided 
\be 
\label{OL9}
\th \neq \alpha \in \R\,,\quad \mbox{or} \quad 
4 \pi n - \th < i \alpha <  4 \pi n + \th \,,\;\; n \in \N_0\,.
\ee
Likewise the constraint (\ref{L8}) is satisfied for 
(\ref{OL5}) only with a divergent constant of proportionality, 
while for (\ref{OL6}) it is satisfied with a finite one. The special 
solution (\ref{OL8}) satisfies (\ref{L8}) with a finite constant of 
proportionality only if formally $\pi < \th < 2 \pi$ is assumed. 
This leaves only (\ref{OL6}) as an acceptable $L^1(\R^2)$ 
solution. The average $\bra \cO \ket_{\varphi_0}$ 
computed from (\ref{OL4a}), (\ref{OL6}) with $\cO(z) = \exp j z$ 
correctly evaluates to 
\be 
\label{OL10}
\bra e^{j z} \ket_{\varphi_0} = 
\exp\Big\{ \frac{e^{i \th/2}}{4 \om} j^2 \Big\} = 
\bra e^{j x} \ket_{\Omega_0}\,,
\ee
where $\Omega_0(x) = e^{-S(x)}$ is the ground state of $\bP$. 
As a final test we also computed the equilibrium distribution 
of the Langevin dynamics numerically and found excellent 
agreement with the analytical formula for $\vp_0$, see 
Figure 7.

\begin{figure}[htb]
\vspace{5mm} 

\begin{center} 
\includegraphics{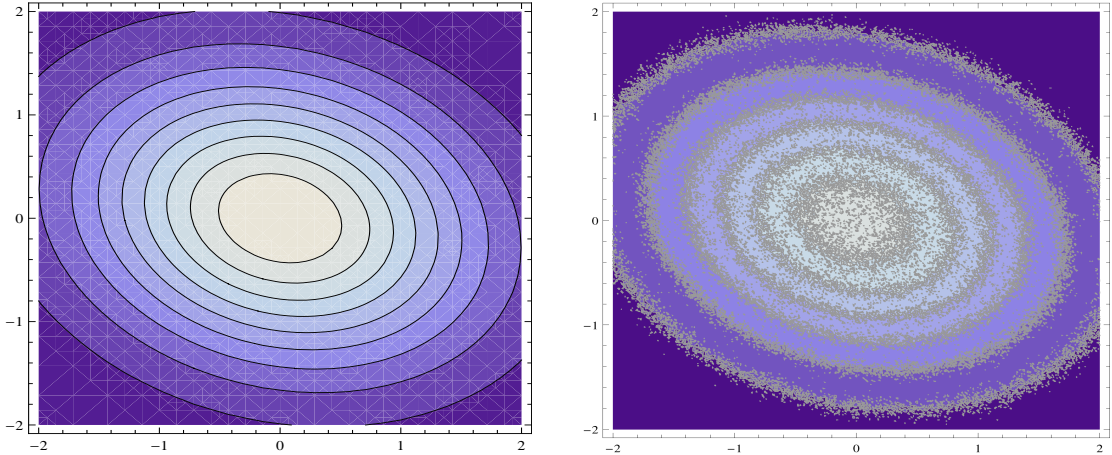}
\end{center} 
\vspace{-6mm}

\caption{\small Ground state wave function for $\P$ with parameters 
$A_R =2, A_I =1, \om=1,\,\th=\pi/2$. Left: contour plot of 
analytical $\vp_0$ from (\ref{OL4a}), (\ref{OL6}). Right: results from
complex Langevin simulations.}
\end{figure}

\newpage
\newappendix{Resummation as a quantum mechanical `supertask'} 
\setcounter{equation}{0}

The recursion relation (\ref{borel2}) for the moments 
is virtually identical to a system of differential equations 
devised by J.~Norton in the context of quantum mechanical `supertasks' 
\cite{Norton}. This gives rise to an amusing parallelism which we 
cannot resist mentioning. In brief, Norton considers a Schr\"{o}dinger 
equation $i \dd_t \psi = H \psi$, where $\psi(t)$ is expanded
with respect to some $L^2$ orthonormal basis
\be 
\label{stask1} 
\psi(t) = \sum_{n \geq 0} f_n(t) e^{i t} \vp_n\,.
\ee
The hamiltonian $H$ is tri-diagonal in the basis 
$(\vp_n)_{n \geq 0}$ and such that the $f_n$ obey the 
recursion relation 
\be 
\label{stask2}
f_n = \frac{a_{n-2}}{a_{n-1}} f_{n-2} - \frac{1}{a_{n-1}} f'_{n-1} \,,
\quad n \geq 2\,,
\ee
with $f_0 = {\rm const}$ and $f_1(t)$ freely specifiable. 
The positive numbers $a_n, \,n \geq 0$, parameterize the 
hamiltonian, see Eqs.~(13') and (17'') in \cite{Norton}. 
In the $n=2$ equation in (\ref{stask2}) we allowed 
$a_0 >0$ for later convenience; compared to $a_0=0$ 
this modifies some aspects related to normalizability 
but does not affect the overall structure. The term `supertask' 
derives from the fact that for suitable choices of $a_n$ 
and initial conditions the system can undergo spontaneous 
excitations with a pattern familiar from classical 
supertasks. That is, each excitation is aroused by a 
faster excitation of higher numbered excitations, ad infinitum, 
leading to severe violations of determinism.   This can 
be achieved without sacrificing quantum mechanical principles,
even in cases where the temporally conserved norm $\Vert \psi \Vert$ 
is finite. The construction capitalizes on the fact 
that the differential form of the time evolution law as set 
by the Schr\"{o}dinger equation $i \dd_t \psi = H \psi$ may 
under-determine the time evolution for systems with an infinite dimensional 
Hilbert space. It is only through the subtle injection 
of additional pieces of information in terms of boundary 
conditions and domains that the associated propagation kernel,
$t \mapsto e^{i t H}$, if well-defined, fully determines the 
time evolution. The countable infinity underlying (\ref{stask2}) 
and the corresponding basis $(\vp_n)_{n \geq 0}$ is essential 
in this context -- any truncation of (\ref{stask2}) at 
some $n \leq N$ would lead to a closed system of differential 
equations which fully determines all functions $f_n, 0 \leq n \leq N$,
including $f_1$. It is therefore a `supertask' to borrow 
information from infinity to specify $f_1(t)$ by some principle.

The case where the $a_n$ grow without bound is referred to as 
an ``accelerated supertask''. In this situation the 
Schr\"{o}dinger equation under-determines the time 
evolution and a wave function with normalizable 
initial condition will in general instantaneously evolve into 
a nonnormalizable one. For example with $f_n(0) = \delta_{n,0}$ 
all $f_n(t)$ are nonzero for arbitrarily small $t>0$, 
with 
\be 
\label{stask3} 
f_n(t) = \frac{1}{n!} a_0 a_1 \ldots a_{n-1} t^n + 
O(t^{n-2}) \,,\quad n \geq 1\,.
\ee 
For $a_n$ growing faster than $n$ this suggests that 
$\sum_{n \geq 0} f_n(t)^2$ will diverge 
for arbitrarily small $t >0$, even if (\ref{stask3}) only
provides the leading terms of an asymptotic expansion.
If the $f_n(t)$ converge for $t \ra \infty$, the limiting 
values are dictated by (\ref{stask1}) 
\be 
\label{stask4} 
f_n(\infty) = \left\{ 
\begin{array}{ll} 
\dfrac{a_0 a_2 \ldots a_{n-2}}{a_1 a_3 \ldots a_{n-1}} f_0 & 
\quad n \;\;\mbox{even} \,,
\\[6mm]
\dfrac{a_1 a_3 \ldots a_{n-2}}{a_2 a_4 \ldots a_{n-1}} f_1(\infty) & 
\quad n \;\;\mbox{odd} \,.
\end{array} \right.
\ee
In simple cases like $a_0 =0, a_n = a, \,n\geq 1$, normalizability 
of the wave function can be used as a criterion to restore uniqueness 
of the time evolution \cite{Norton}. In the situation (\ref{stask3}), 
(\ref{stask4}) above normalizability is necessarily violated when 
$a_0 f_1(\infty) \neq 0$. To see this note that 
\be
\label{stask5} 
\Vert \psi \Vert^2 := \sum_{n \geq 0} f_n(t)^2 - 2 a_0 f_0 \int_0^{t} \!ds
\, f_1(s) \,,
\ee
is formally conserved on account of (\ref{stask2}). The rearranging 
of the sums is legitimate only if $\sum_{n \geq 1} a_n f_n(t) f_{n+1}(t)$ 
converges absolutely for all $t$. In the $t \ra \infty$ limit the 
latter sum contains denumerably many identical terms of the form 
$f_0 a_0 f_1(\infty)$; so with $f_0 \neq 0$ the vanishing of 
$a_0 f_1(\infty)$ is a necessary condition for normalizability. 
In the following we are interested in situations where 
$a_0 f_1(\infty) \neq 0$ and argue that, instead, Borel summability 
can be used as a criterion to restore uniqueness of the time 
evolution. To this end we make contact with the results of Section 3.2.

Consider specifically the case 
\be 
\label{stask6} 
a_n = \sqrt{2 n (2 n +1) (2 n +2)} \,, \quad n\geq 1\,,
\ee
and set $m_0 =1$ and
\be 
\label{stask7} 
m_p(t) = (4 \alpha)^{-p/4} 
\frac{\sqrt{2p (p-1)!!}}{a_0 f_0} 
f_{p/2}(\sqrt{4 \alpha} t) \,, \quad p=2\!\! \mod 2\,.
\ee
Then the recursion relation (\ref{borel2}) for the moments 
is mapped into (\ref{stask2}) with the specific $a_n$ in (\ref{stask6}) 
and $f_0=1$. In particular (\ref{stask3}) is mapped into 
$m_p(t) = \frac{p!}{(p/2)!} t^{p/2} + O(t^{p/2 -1})$, as required,
and the asymptotics (\ref{stask4}) is mapped onto (\ref{QL9}). 
As seen in Section 3.2 the formal power series in $t$ 
obtained from (\ref{borel2}) is asymptotic to the exact result 
for short times. It is also 
Borel summable and the Borel transform uniquely defines 
an $m_2(t)$ for all times with a nonzero asymptotics $m_2(\infty)$. 
Hence all moments $m_p(t)$ are uniquely determined and 
as shown in Section 3 they also coincide with the ones
defined by the complex propagation kernel via (\ref{rhotrans}).

Hence the Borel transform in this case augments precisely the 
piece of information otherwise supplied by the construction 
of the propagation kernel, the latter however cannot be 
achieved explicitly even in the simple case of a sextic 
anharmonic oscillator. The failed equivalence of (\ref{rhotrans}) 
to the $\bra z^p\ket_{R_t}$  moments highlights that a differently 
constructed semigroup may define a different time evolution 
with a different $t \ra \infty$ asymptotics. 

Although (\ref{stask6}), (\ref{stask7}) provides a mathematical 
isomorphism the conceptual interpretation of the $m_p$ and the $f_n$ 
is of course different. The $m_p$ are averages of the observables 
$x^p$ and are independent of the choice of basis in the underlying 
state space. Their dynamics (\ref{borel2}) arises from instances 
of the observable flow equation (\ref{oflow2}) with respect to the 
stochastic time $t$.  The dynamics of each fixed observable is driven by 
the quartic action  via $\bL = - e^{S/2} \bH e^{-S/2} = 
\dd_x^2 - \dd_x S \dd_x$, and only when a complete set of observables 
is considered is the effective dynamics of the set 
governed by the tri-diagonal hamiltonian underlying (\ref{stask2}). 
The $f_n$ in (\ref{stask1}), on the other hand, are the coefficients 
of the time dependent Schr\"{o}dinger wave function with respect 
to a preferred basis, and as such do not directly qualify as 
observables. The time variable $t$ refers to the physical time. 
Nevertheless the correspondence (\ref{stask7}) cuts both ways. 
It could be used to design examples where the quantum mechanical 
supertasks can be solved by Borel transform. Conversely the 
under-determination in the Schr\"{o}dinger dynamics (\ref{borel2}) 
is ultimately the reason for the failure of the Parisi-Klauder 
conjecture in the present context.   

\newpage 


\end{document}